\documentclass{aa}
\usepackage{graphicx}
\usepackage{longtable}
\usepackage{latexsym}
\usepackage{amssymb}
\usepackage{lscape}
\begin{document}
   \title{The structure of elliptical galaxies in the Virgo cluster. Results from the INT Wide Field Survey}

   \author{G. Gavazzi\inst{1}, A. Donati\inst{1}, O. Cucciati \inst{1}, S. Sabatini\inst{2}, A. Boselli\inst{3}, J. Davies\inst{2},
   S. Zibetti\inst{4}}

   \offprints{G.Gavazzi}

   \institute{Universit\'{a} degli Studi di Milano-Bicocca, P.zza della Scienza 3, 20126 Milano, Italy.\\
              \email{Giuseppe.Gavazzi@mib.infn.it; alessandro.donati@mib.infn.it}
          \and
             Department of Physics and Astronomy, Cardiff University,\\
             \email{sabina.sabatini@astro.cf.ac.uk}      
          \and
             Laboratoire d'Astrophysique de Marseille, BP8, Traverse du Siphon, F-13376 Marseille, France.\\
             \email{alessandro.boselli@oamp.fr} 
	  \and
             Max-Planck-Institut für Astrophysik, Garching, Germany.\\
             \email{zibetti@MPA-garching.MPG.DE}            
             }

\date{Received 24/10/2003; accepted 21/9/2004.}

\abstract{
We report on a complete CCD imaging survey of 226 elliptical galaxies in the North-East 
quadrant of the Virgo cluster, representative of the properties of giant and dwarf elliptical
galaxies in this cluster. We fit their radial light profiles with the Sersic $r^{1/n}$ model of light distribution.
We confirm the result of Graham \& Guzman (2003) that the apparent dichotomy between 
E and dE galaxies in the luminosity--$<\mu>_e$ plane no longer appears when other structural parameters
are considered and can be entirely attributed to the onset of "core" galaxies
at $B_T \sim -20.5$ mag. When "core" galaxies are not considered,
E and dE form a unique family with $n$ linearly increasing with the luminosity. \\
For 90 galaxies we analyze the B-I color indices, both in the nuclear and in the outer regions.
Both indices are bluer toward fainter luminosities. 
We find also that the outer color gradients do not show any significant correlation with the luminosity.
The scatter in all color indicators increases significantly toward lower luminosities, e.g.
galaxies fainter than $B_T \sim -15$ have a B-I spread $>$0.5 mag.

}

\titlerunning{INT WFS in Virgo}
\authorrunning{G. Gavazzi et al.}

   \maketitle
%

\section{Introduction}

The widespread belief that dwarf (dE) and giant (E) elliptical galaxies form two distinct families
brought unanimous consent that separate mechanisms are responsible for their formation and evolution
(see Ferguson \& Binggeli 1994 for a review).
Alleged evidence for a structural dE/E dichotomy includes: 
i) the light profiles of dEs follow exponential 
laws (Binggeli et al. 1984), as opposed to their giant counterparts 
that follow the $r^{1/4}$ law (de Vaucouleurs 1948); 
ii) scaling relations of the structural parameters 
$r_e$, the effective radius, and $<\mu>_e$, the mean surface brightness within $r_e$, 
with the total luminosity indicate that dEs of increasing luminosity have greater 
surface brightness while giant Es show the reverse trend (Binggeli \& Cameron 1991, Binggeli et al. 1984). 
Moreover they occupy distinct loci 
in the $r_e$ vs. $<\mu>_e$ plane: giants have dimmer surface brightness with increasing
radius (Kormendy \& Djorgowski 1989), while dEs do not.\\
Hydrodynamical simulations, including mass loss from stars and 
gas heating and cooling, show that gravitational instability of primeval density fluctuations 
(with or without dark matter) results in the formation 
of elliptical galaxies with brighter $<\mu>_e$
at high luminosity, as observed in the dE regime, without showing any dichotomy, i.e.
the bright objects result as scaled-up versions of the less luminous systems
(Athanassoula 1993). \\
Various mechanisms involving merging have been proposed for the formation of E galaxies
that are consistent with the alleged dE/E dichotomy. 
Dwarf Es in clusters might derive from the harassment of dwarf spiral galaxies (Moore 
et al. 1996; Moore, Lake, \& Katz 1998; Mao \& Mo 1998; Mayer et al.
2001), while giant E galaxies might arise from the merger of two spiral galaxies  
or from the multiple merging of dEs (Kauffmann et al. 1993).\\
There is however some observational evidence arguing for a continuity rather than a dichotomy between dEs and Es:
i) the U-V (Caldwell 1983) and B-H (Scodeggio et al. 2002) colors and the metallicity are found to smoothly increase with 
luminosity; ii) the Near-IR light concentration index (Scodeggio et al. 2002) and the central 
surface brightness (Caldwell \& Bothun 1987; Jerjen \& Binggeli 1997) increase monotonously with the absolute magnitude;
iii) no clear dichotomy can be assessed between giants and dwarfs 
in the Near-IR $<\mu>_e-R_e$ and fundamental plane relations (Zibetti et al. 2002). 
\begin{figure*}
\centering
\includegraphics[width=16.0cm]{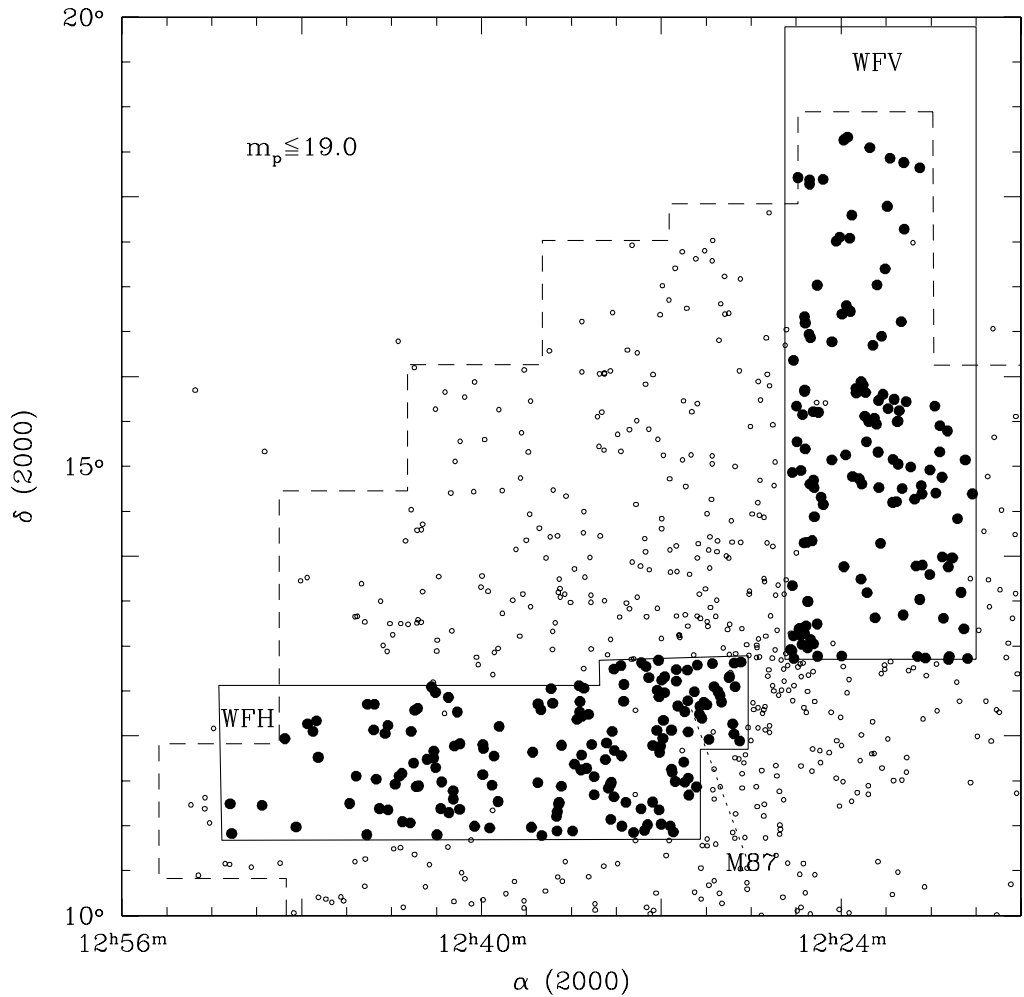}
\caption{The N-E region of the Virgo cluster including galaxies with $m_p\leq19.0$. The dashed line represents
the boundary of the VCC catalog. The rectangles include the studied WFH and WFV areas, 
covering approximately 26 $deg^2$. The location of M87 is indicated.}
\label{sample}
\end{figure*}
Whether there is kinematic evidence for the dE/E dichotomy
remains a controversial issue: the preliminary conclusion of Geha et al. (2001) that
dE and E are both not rotation-supported systems has been subsequently questioned by the
same authors (Geha et al. 2003) who concluded, in agreement with Pedraz et al. (2002) and van Zee et al. (2004), 
that dE galaxies can have significant rotation.\\
Moreover the very existence of the aforementioned dE/E dichotomy has been recently questioned by 
Graham \& Guzman (2003; GG03 hereafter).
These authors argued that the dichotomy is apparent, as it derives from the use of 
exponential and de Vaucouleurs laws to decompose the light profiles of dE and E galaxies respectively, 
combined with the use of $<\mu>_e$ as representative of the surface brightness. When $\mu_o$ is analyzed,
as obtained from the inner extrapolation of Sersic $r^{1/n}$ laws (Sersic 1968), no dichotomy remains.\\
Stimulated by the conclusions of GG03
we undertook this project aimed at re-assessing the relations involving the size, the
surface brightness and the luminosity of dE-E galaxies, all ingredients entering in the determination of the 
fundamental plane. The present analysis is carried out using a sample    
which by selection should be representative of the properties of dE-E galaxies in one cluster.\\ 
The Virgo cluster, owing to its relatively small distance from us (17 Mpc), represents  an
appropriate testbed for such a study, although its significant line-of-sight depth (Gavazzi et al. 1999; 
Solanes et al. 2002) and sub-clustering makes it not ideal.
In spite of their low surface brightness, dwarf galaxies as faint as 19 mag 
are within reach of mid-size telescopes and seeing limitations are less severe when studying galaxies
at the distance of Virgo than objects 5 times further away such as the Coma galaxies 
that require the superior resolution of HST. 
Since their luminosity function was determined ( e.g. Sandage et al. 1985), early 
systematic studies of dwarf galaxies in Virgo were based on photographic material 
(Binggeli \& Cameron 1991, 1993,  Young \& Currie (1998)).  
Modern photometry obtained with panoramic detectors exists today for many Virgo galaxies
(e.g. Caon et al. 1993; Gavazzi et al. 2003; Barazza et al. 2003) and many have been observed with the HST 
(e.g. Faber et al. 1997; Cote et al. 2004).  
A large stretch of the Virgo cluster was mosaiced
with the Isaac Newton Telescope in what is known as the 
"Virgo Wide Field Survey" (Sabatini et al. 2003).
These data are used in the present analysis to re-assess the issue of the dE/E dichotomy.\\ 
The analyzed sample is described in some detail in Section 2.  Sections 3 and 4 describe
the method used to derive the light profiles and their fitting with Sersic models.
The B-I color analysis and the analysis of the various structural 
B-band parameters are carried out in Section 5 and the
conclusions of our work are briefly summarized in Section 6.

\section{The sample}

\begin{table}  
\caption{The sample. For each band the number of dE-E galaxies in the WFS (separately for the
Horizontal and Vertical strips) with $m_p\leq 19$ mag.
that are members or possible members, to which we have been able to fit a Sersic profile is given. The
denominator gives the number of galaxies in the VCC at the same magnitude limit.} 
\label{Tsample} \[ 
\begin{tabular}{ccc} 
\hline 
\noalign{\smallskip}
      &   B  & I \\
Strip &    $m_p\leq19$  & $m_p\leq19$ \\      	
\noalign{\smallskip} 
\hline 
\noalign{\smallskip} 
WFH & 131/149 (88\%) &     -   \\
WFV &  95/102 (93\%) & 90/102 (89\%) \\
\noalign{\smallskip} 
\hline 
\end{tabular} \] 
\end{table} 
The present structural analysis of elliptical galaxies is focused on the North-East 
quadrant of the Virgo cluster delimited
by $\alpha>12^h20^m00^s$ and $\delta>10^o00'00"$ (see Fig.\ref{sample}).
With this choice we try to include in the analysis genuine members of cluster A (M87) and to exclude members of
other dynamical units within the Virgo cluster, namely cluster B (M49), clouds M and W which are
located to the West and South of the above region, showing significant 3-D structure, i.e.
distances in excess of 6 (cluster B) and 15 Mpc (M, W clouds) with respect to cluster A that we assume at 17
Mpc distance (Gavazzi et al. 1999).\\
\begin{figure}
\centering
\includegraphics[width=9.0cm]{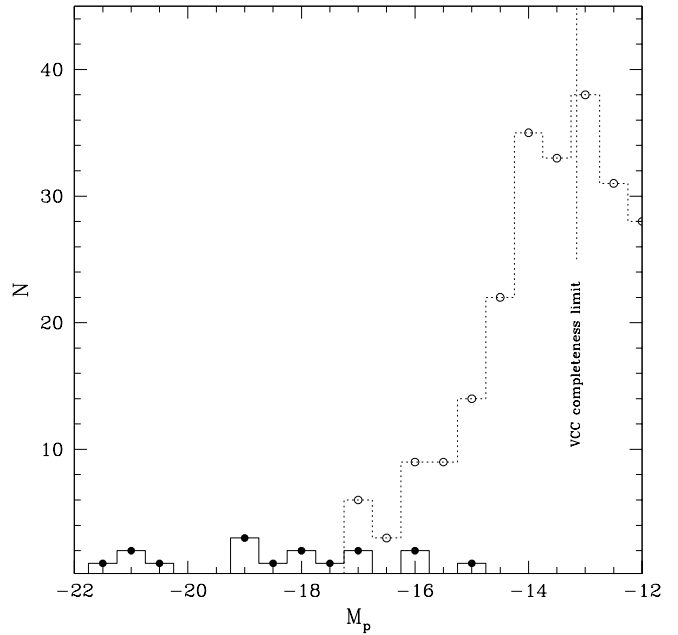}
\caption{The luminosity function of E (filled symbols) and dE (open symbols) in the WFS.
The completeness limit of the VCC ($m_p=18$) is indicated.}
\label{lumfun}
\end{figure}
Our structural analysis requires the availability of imaging material 
(CCD images) suitable for deriving the light profiles according to the method described in Section 4. 
The imaging data are taken from the Wide Field Survey (WFS), horizontal (WFH) and vertical (WFV) strips 
delimited by the solid rectangles in Fig.\ref{sample}. B-band data are available for both strips.  
Moreover in the vertical strip we had access to I-band data to study the B-I color of galaxies
in this region (see Section \ref{colorIB}). 
WFS images are in principle
available for all (439) $m_p\leq 20$ galaxies in the Virgo Cluster Catalog (VCC; Binggeli et al. 1985) 
projected onto the WFS region.\footnote{Nine VCC galaxies in the WFH 
(VCC 1143, 1147, 1216, 1335, 1336, 1482, 1486, 1795, 1886) and 
nine in the WFV (281, 333, 400, 437, 723, 742, 769, 863, 884) were not observed in the B-band 
because they lie in the gaps between adjacent chips of the WFC; similarly 
15 objects in the WFV (I-band) (VCC 400, 413, 505, 578, 663, 666, 677, 723, 751, 769, 797. 798. 857, 882, 884).}
We restricted however the present analysis to the 251 galaxies with $m_p \leq 19$, with morphological type dE or E, that are
spectroscopically confirmed members ($\rm V<3000 ~km ~s^{-1}$) or possible members according to 
Binggeli et al. (1985) and Binggeli, Popescu \& Tammann (1993). We were able to fit meaningful B Sersic profiles to 226 galaxies out of 251 because 
18 were not observed and the remaining 7 were too faint or the images were of insufficient quality. 
Summarizing, B-band structural information is available for 226/251 objects, thus with a completeness of 90\%.
The I-band structural information is available for 90 objects.
The sample completeness is detailed in Table \ref{Tsample}. 
At the assumed distance of 17 Mpc ($\mu=-31.15$), the 
limit of $m_p = 19$ mag corresponds to $M_p$=-12.15. 
Therefore the analyzed sample can be considered representative 
of the properties of dE-E galaxies in the N-W portion of the Virgo cluster.
Fig. \ref{lumfun} illustrates the luminosity function of dE-E galaxies
included in the present analysis. It is in full agreement with the 
one derived by Sandage et al. (1985, see their Fig. 11) for the entire VCC, once
re-normalized to account for the 28\% coverage of the WFS with respect to the entire cluster
and for the different distance modulus assumed.

\section{The imaging data}

\begin{figure}
\centering
\includegraphics[width=9.0cm]{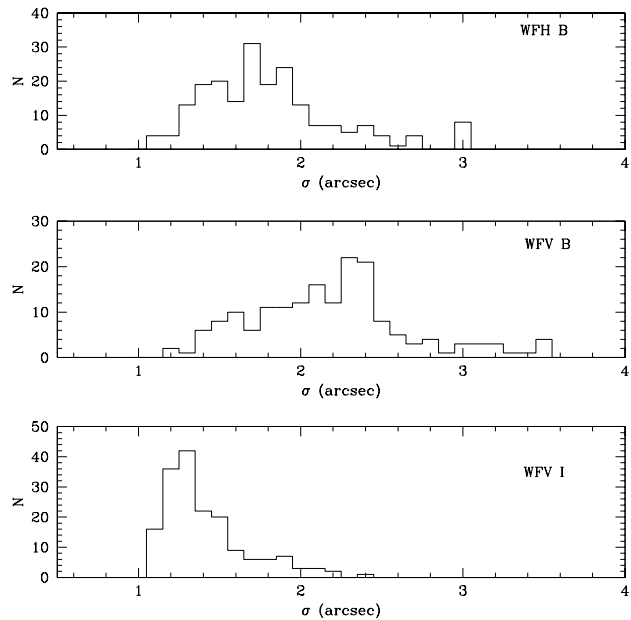}
\caption{Distribution of the seeing in the WFS.}
\label{figseeing}
\end{figure}

The Virgo Wide Field Survey was carried out
with the Wide Field Camera (WFC) at the Isaac Newton Telescope (INT) (Sabatini et al. 2003).\\
The horizontal strip was covered with 55 overlapping fields 
of 34x22 arcmin each, using 3 of the 4 chips of 4000x2000 pixel (each of 0.333 arcsec on the sky) 
constituting the WFC\footnote{The WFC is made of 3 chips, one on top of the other,
and of one orthogonal chip on one side. The top horizontal chip is partly vignetted. This chip was not used
in the B-band survey, but it was considered in the I-band (see next Section).}. 
For this strip we had access only to the B-band (Johnson) data.
The observations were carried out in nearly photometric conditions with a seeing 
varying from 1 to 3 arcsec, with a mean of 1.6 arcsec (see Fig.\ref{figseeing}). An integration time 
of 750 seconds was used.\\
The Vertical strip was covered in both the B- and the I-band. The B-band survey was carried out
with 80 overlapping fields, using 3 of the 4 chips of the WFC, each exposed for 750 seconds.
The observations were taken in nearly photometric conditions with a seeing 
varying from 1.2 to 3.5 arcsec, with a mean of 2.1 arcsec.\\
The I-band survey consists of
73 overlapping fields\footnote{The 7 missing I-band frames lie at $\delta>18^o00'$, in a region
almost devoid of VCC galaxies.}, 
each exposed for 1000 seconds.
The observations were carried out in nearly photometric conditions with a seeing 
varying from 1 to 2.5 arcsec, with a mean of 1.3 arcsec.\\

\section{Data reduction}

The reduction of the B-band science frames was performed as described in Sabatini et al. (2003).
The zero point of all images was obtained from the "CASU INT Wide Field Survey Home Page"
(http://www.ast.cam.ac.uk/$\sim$wfcsur/index.php).\\ 
The I-band images suffer from fringing, producing
unwanted structures in the background. However for this band we could access images produced 
by all 4 chips of the WFC, including the top, redundant, vignetted chip. 
Several galaxies were observed in non-vignetted parts of this chip as well as
in other chips of other frames. In these cases we combined two independent sets of measurements
to reduce the fringing.\\
Because of the significantly better seeing in the I- compared to the B-band, 
the I-band profile decomposition (see next section) was performed twice: once on the
original data, once on degraded images obtained convolving the images to the B-band seeing.

\subsection {Profile decomposition.}

The profile decomposition is performed using the method described in
Gavazzi et al. (2000, 2001) which is based on 
the IRAF environment and relies on the STSDAS 
\footnote{IRAF is the Image Analysis and Reduction Facility made
available to the astronomical community by the National Optical
Astronomy Observatories, which are operated by AURA, Inc., under
contract with the U.S. National Science Foundation. STSDAS is
distributed by the Space Telescope Science Institute, which is
operated by the Association of Universities for Research in Astronomy
(AURA), Inc., under NASA contract NAS 5--26555.} package and on GALPHOT
(developed for IRAF- STSDAS mainly by W. Freudling, J. Salzer, and
M.P. Haynes and adapted by M. Scodeggio, P. Franzetti and S. Zibetti).\\ 
For each frame the sky background is determined as the mean number of
counts measured in regions of ``empty'' sky, and it is subtracted
from the frame.\\ 
The 2-dimensional light distribution of each galaxy is fitted with
elliptical isophotes, using a procedure based on the task ${\it
ellipse}$, (STSDAS ${\it ISOPHOTE}$ package; Jedrzejewski 1987, and
Busko 1996), which allows the interactive masking of unwanted
superposed stars and galaxies.  Starting from an interactively centered ellipse,
the fit maintains as free parameters the ellipse center, ellipticity
and position angle. The ellipse semi-major axis is incremented by a
fixed fraction of its value at each step of the fitting procedure.  The
routine halts when the surface brightness found in a given annulus equals
the the average sky value, and then restarts decrementing the initial semi-major axis
toward the center.  
\begin{figure}
\centering
\includegraphics[width=9cm]{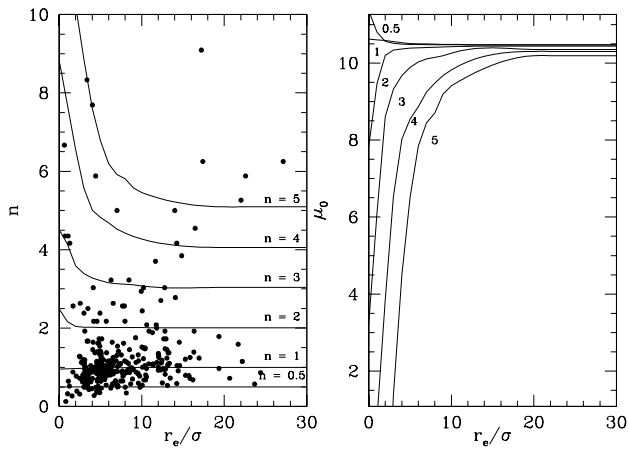}
\caption{The effect of the seeing on $n$ (left) and on $\mu_0$ (right) as derived from the simulations (lines)
for various $n$. Dots give the distribution of the observed points in the $n$ - $r_e/\sigma$ plane.}
\label{seeing_corr}
\end{figure}
The fit fails to converge for some very faint galaxies. In these cases we keep fixed one or more
of the ellipse parameters.\\ 
The resulting surface brightness profiles are fitted using the Sersic model
(Sersic 1968) of light distribution: 
   \begin{eqnarray}
   \lefteqn{I(r) = I_0 e^{-(r/r_o)^{1/n}}}
   \end{eqnarray}
where $I$ is the surface brightness (in intensity) at the 
radius $r$ (along the major axis), $I_0$ is the 
central surface brightness, $r_0$ the scale length and $n$ is the dimensionless shape parameter that 
determines the curvature of the profile.
This model is a simple generalization of de Vaucouleurs ($n=4$) and exponential ($n=1$) law.
The surface brightness profiles are fitted with this model in the magnitude representation:\\
\begin{eqnarray}
\lefteqn{\mu(r) = \mu_0 + 1.086 (r/r_o)^{1/n}}
\end{eqnarray} 
where $\mu(r)$ is the surface brightness (in magnitude) at the radius $r$ (along the major axis), 
$\mu_0$ is the central surface brightness.\\
The fit is performed using
a weighted least squares method  
from a radius equal to one seeing disk ($\sigma$)(with the exceptions discussed in Sect.\ref{seeing}), out
to the outermost significant isophotes, i.e. when the surface brightness equals
the average sky value. We deliberately avoid fitting the nuclear features.\\
The total asymptotic magnitude $B_T=-2.5log(F_T)$ is obtained by adding to the flux measured ($F_{last}$)
within the outermost significant isophote ($r_{last}$) the extra flux extrapolated to
infinity along the model:
\begin{eqnarray}
\lefteqn{F_{extra} = 2\pi  I_0   (1-\epsilon)  n  r_o^2 ~\big[\Gamma(2n)-\gamma\big(2n,\big(\frac{r_{last}} {r_0}\big)^{1/n}\big)\big]}
\end{eqnarray}
where $\epsilon$ is the ellipse eccentricity, $\Gamma$ and $\gamma$ are the complete and incomplete
Gamma functions respectively.
The effective radius $r_e$ (the radius containing half of the total
light) and the effective surface brightness $<\mu>_e$ (the mean surface
brightness within $r_e$), $r_{25}$, $r_{75}$ (the radii that enclose 25\% and 75\% of the total light) 
of each galaxy are computed empirically from $B_T$, i.e. are derived from the data, using
the Sersic models to extrapolate from the last measured magnitude to infinity.
Similarly we compute the concentration index
($C_{31}$), defined in de Vaucouleurs (1977) as the
ratio between $r_{75}$ and $r_{25}$.
Fits to the B-band light profiles were obtained for 136 galaxies, as given in Fig. \ref{profiles},
and to B+I-band light profiles for 90 objects, as given in Fig. \ref{profilesBI}.

\subsection{The effects of the seeing}
\label{seeing}

Attempts to model the effects of the seeing on Sersic parameters
can be found in the literature, e.g. by Trujillo et al. (2001a,b) who modeled the effect of Gaussian and Moffat
convolution on Sersic profiles. To illustrate this issue we used some simple simulations.
We constructed sets of artificial images injecting fake Sersic galaxies with zero ellipticity, 
constant $\mu_0$, fixed $r_e$=10 arcsec and Sersic index $n$ varying
from 0.5 to 5 on a sky background with noise characteristics similar to the INT frames. 
To simulate the effects of the seeing the images were convolved with Gaussians  
of FWHM ($\sigma$) ranging from 0.5 to 10 arcsec, so that $r_e/\sigma$ varies between 20 and 1, mimicking
the range covered by the real data.
We fitted circular isophotes and measured the Sersic parameters on the blurred images with
the same tools and criteria used for real data (i.e. disregarding the data points
within the seeing disk). 
Figure \ref{seeing_corr} illustrates how the measured Sersic parameters $n$ and $\mu_0$ 
vary as a function of $r_e/\sigma$ (lines). Dots give the observed distribution in the
$n$--$r_e/\sigma$  plane.
From Fig. \ref{seeing_corr} it is apparent that increasing 
corrections to the Sersic parameters are required with increasing $n$ and decreasing $r_e/\sigma$.
For $n\geq3$ the seeing produces overestimates of $n$ (thus enhancing the central 
surface brightness $\mu_0$) because flux from the central cusp is distributed
in the surrounding pixels producing a steepening of the inner profile.\\
As a result of the simulations we decided not to try to correct the individual Sersic parameters
(that are not independent of one another), but to modify our fitting strategy in the case of $n\geq3$,
i.e. excluding from the fit the data points within $2\times$ the seeing disk $\sigma$.
\begin{figure}
\centering
\includegraphics[width=7.5cm]{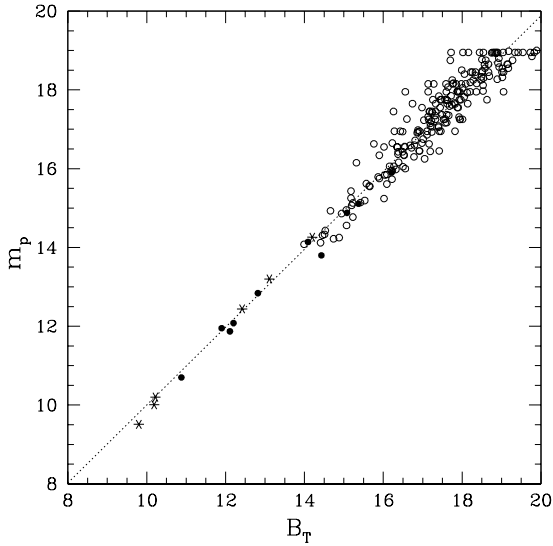}
\caption{Comparison of $B_T$ from this work with $m_p$  from the VCC. The dotted line
gives the bisector linear regression. In this and in the following figures (unless otherwhise specified)
empty symbols refer to dEs, filled symbols to Es and asterisks to "core" galaxies.}
\label{mag}
\end{figure} 
\begin{figure}[t]
\centering
\includegraphics[width=9cm]{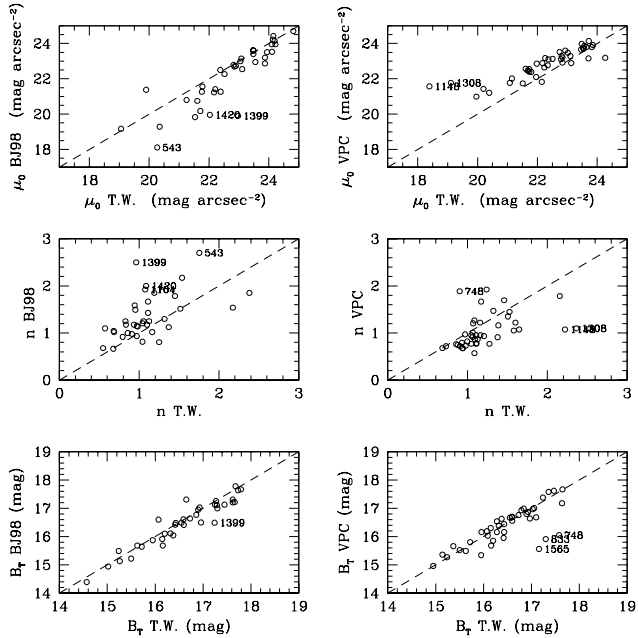}
\caption{Comparison among structural parameters derived in this work and by Young \& Currie (1998) for 42 common galaxies (right) and
by Binggeli \& Jerijn (1998) for 37 common galaxies (left).}
\label{comp}
\end{figure}
\begin{figure}[b]
\centering
\includegraphics[width=9.0cm]{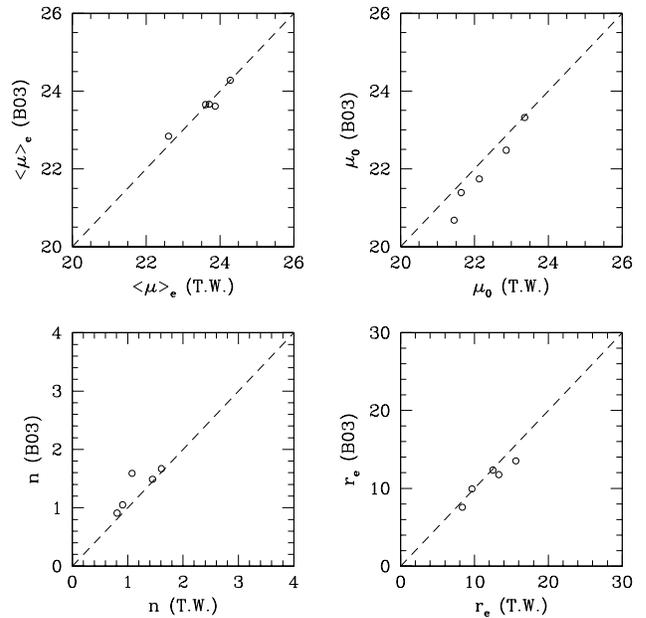} 
\caption{Comparison among structural parameters derived in this work and by Barazza et al. (2003) for 5 common galaxies.}
\label{barazza}
\end{figure}

\section{Results}

The  B-band parameters derived in this work are listed in Table 2.
Magnitudes and surface brightness are corrected 
for extinction in our Galaxy according to Burstein \& Heiles (1982)
(due to the high galactic latitude of the Virgo cluster
this results in a small 0.05 mag correction on average included in Columns 4, 6, 7, 9).

\subsection{Consistency test}

The quality of the measurements in Table 2 is first checked by comparing the total asymptotic 
magnitudes ($B_T$) derived in
this work with $m_p$ given in the VCC (see Fig. \ref{mag}).
$B_T$ is found in linear proportionality with $m_p$. The bisector linear regression
(see Feigelson \& Babu 1992) is:\\
$m_p$ = +0.11 + 0.99 $B_T$ (R=0.96)\\
where R is the Pearson regression coefficient.\\
Fig. \ref{comp} illustrates the comparison between the parameters derived in this work on CCD frames
with the parameters derived on photographic material
by Young \& Currie (1998, VPC) for the 42 common galaxies (left) and by Binggeli \& Jerjen (1998) (based on 
photographic data of Binggeli \& Cameron 1991) for the 37 common galaxies (right). 
The agreement between the two sets of measurements is satisfactory, except for some deviating 
objects marked individually
in the figure with their VCC denominations. The three most discrepant objects 
(i.e. VCC 748, 833, 1565) have $B_T$  derived in this work consistent within 0.28 mag with $m_p$ from the VCC,
differing by more than 1.5 mag from the VPC values. Furthermore we checked the $n$ index on the profiles of 
VCC 748, 1148 and 1308 and confirmed that the value found in this work is correct.\\
We remark that 18 galaxies are in common between this work, Young \& Currie (1998) and Binggeli \& Jerjen (1998).
For these objects we compare the Sersic parameters on 3 independent data-sets and we conclude that
the consistency is best between this work and the VPC and worst between Young \& Currie and Binggeli \& Jerjen (1998).
In other words the average errors in the VPC are approx. 1.5 times larger than those in this work while 
those of Binggeli \& Jerjen (1998) are approx. 3 times larger than those in this work.\\
Indeed when we compare the results of this work with those obtained by Barazza et al. (2003) 
on CCD material for the 5 common galaxies (see Figure \ref{barazza}) we find a high consistency.

\subsection{B-I color analysis}
\label{colorIB}

\begin{figure}
\centering
\includegraphics[width=6.5cm]{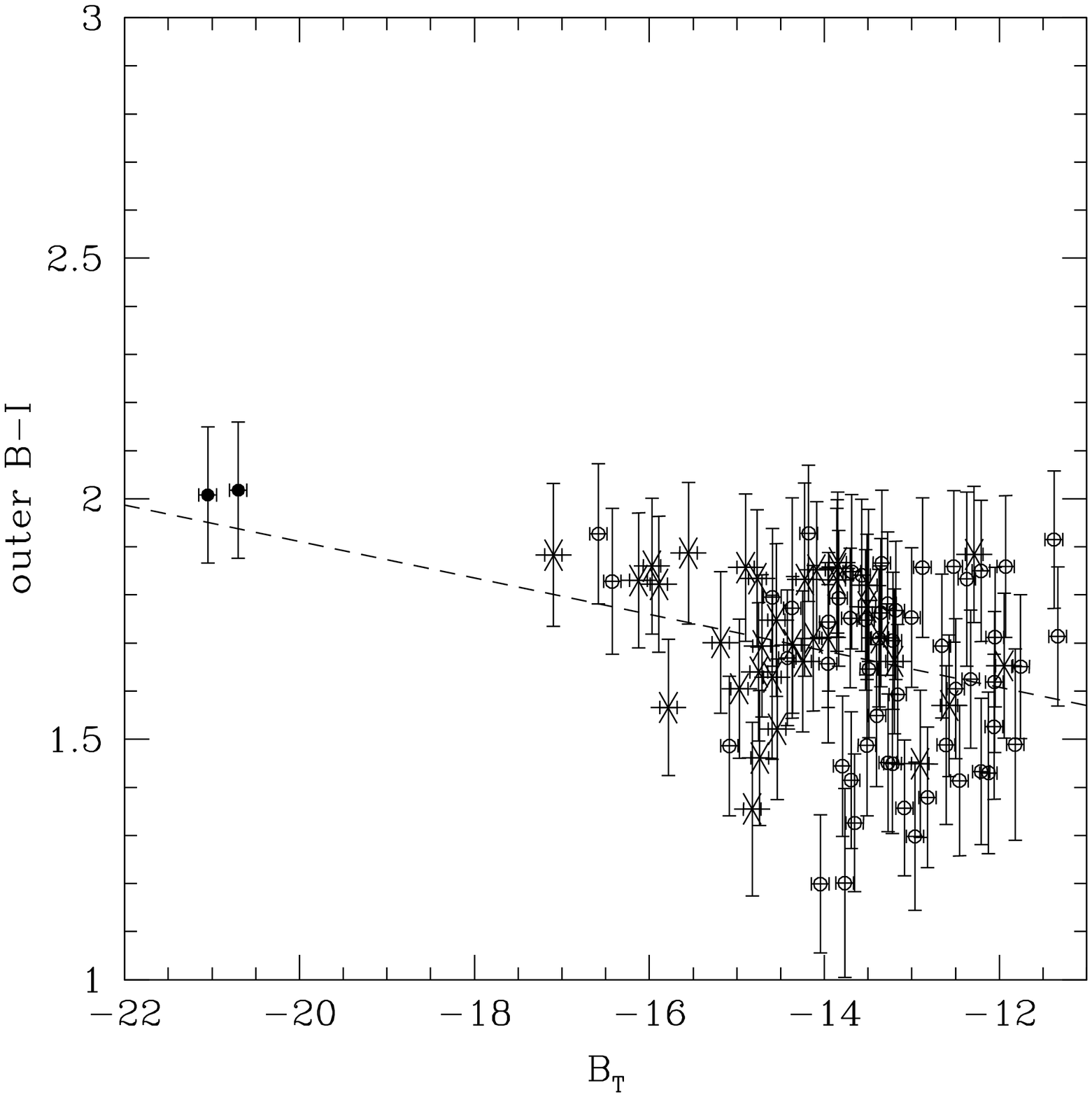}
\includegraphics[width=6.5cm]{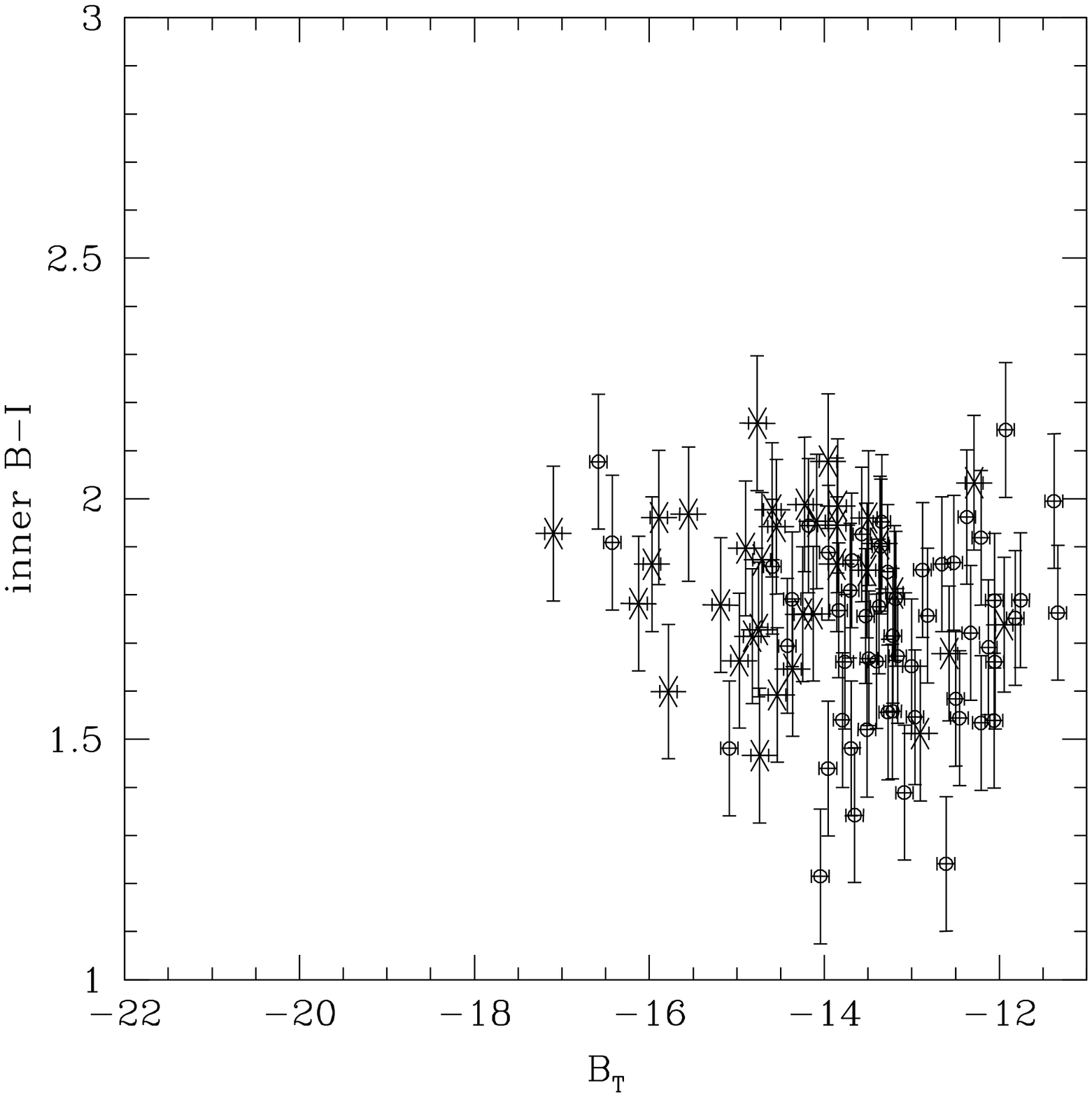}
\includegraphics[width=6.5cm]{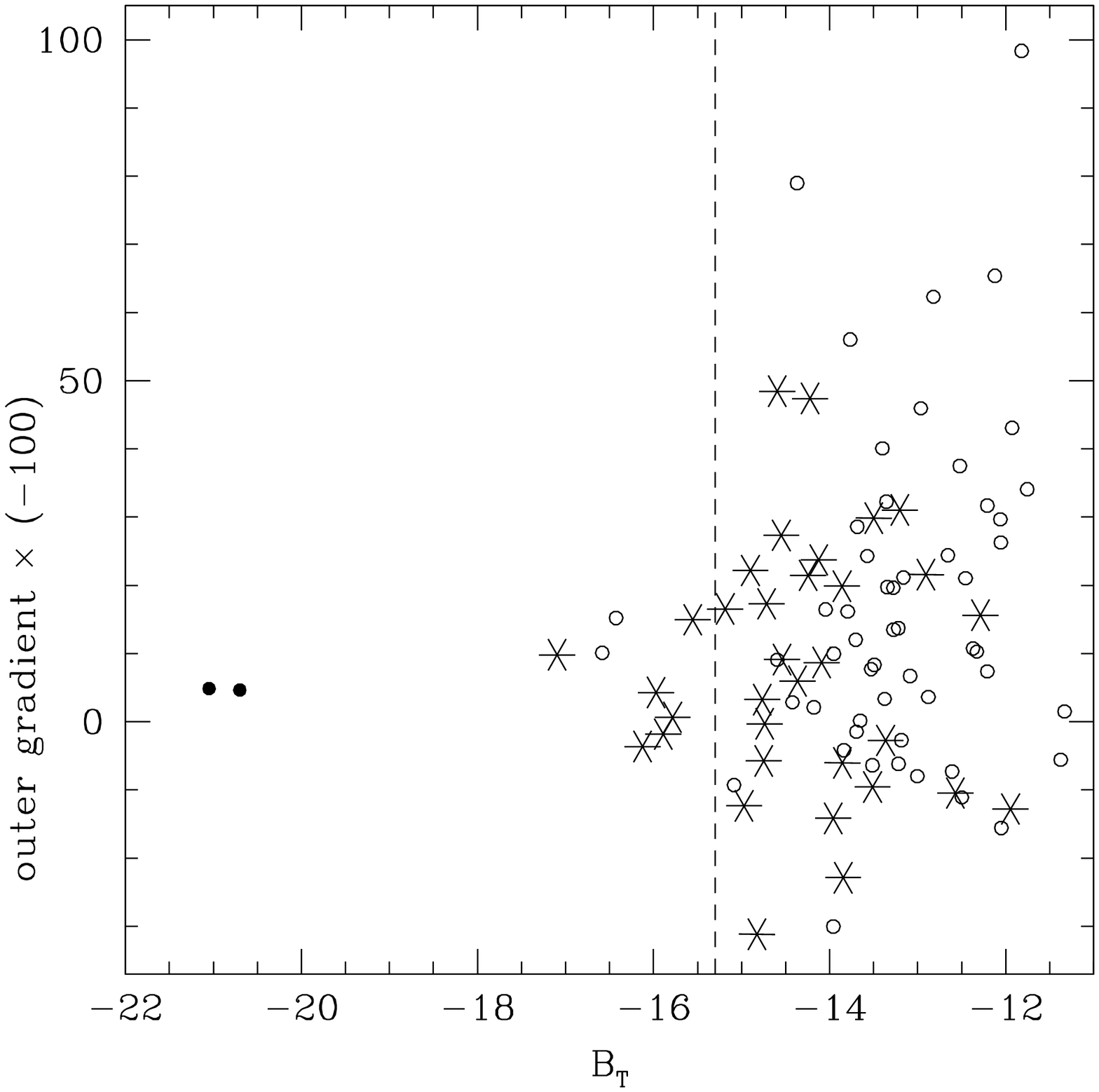}
\caption{The outer B-I color index  (top); the inner B-I color index (middle; the two brightest objects are missing
in this panel because they are both saturated) and 
the outer B-I gradient (bottom) as a function of $B_T$. In this figure filled dots represent giant Es, empty dots dEs, 
asterisks Nucleated dEs.}
\label{colorB-I}
\end{figure}
Using 90 galaxies in the WFV that have been observed both in the B and in the I-band 
we analyze in this section the B-I color properties of dE-E galaxies.
For these objects the B band Sersic parameters were compared with two sets of similar I band parameters,
i.e. those derived from the original I images and from the I images convolved to the B-band seeing.
The agreement between the B and I parameters is satisfactory and it improves when
the B parameters are compared with those obtained from the blurred I images.\\
For these objects we also compute the outer B-I color index in the radial interval between
$r>1.5 \times \sigma$ and (two points before) the last significant measurement, as representative
of the galaxy color outside the nucleus, avoiding the least reliable, outermost points. 
Figure \ref{colorB-I} (top) gives the color-magnitude diagram obtained in this way, 
showing the expected 
trend of blueing toward fainter luminosities (see Baldry et al. 2004).  We remark however that our WFV data-set
contains too few giant E galaxies to explore the relation at high luminosity.
It appears that, fainter than $B_T=-16$, the scatter in the relation becames increasingly large.\\ 
For the same set of galaxies we also compute the "outer color gradient"
as the slope of the line fitted to the B-I color profile
in the radial interval between
$r>1.5 \times \sigma$ and (two points before) the last significant measurement.
When plotting the outer gradient we follow the convention of Vader et al. (1988) of "positive" gradients
when the the center is redder.
Figure \ref{colorB-I} (bottom) shows the outer gradients as a function of the B luminosity, revealing
no significant correlation. We do not confirm the trend found by 
Vader et al. (1988) in a brighter ($B_T<-15.3$, vertical line) sample 
who claimed that brighter galaxies have more positive gradients (outer envelopes bluer than inner regions) than 
fainter objects.
Our fainter sample shows that this trend does not hold at faint luminosities where
the outer gradients show a tremendous scatter.\\  
Figure \ref{colorB-I} (middle) shows the inner (or nuclear) color index obtained as the difference
of the B and I magnitudes integrated within  $1 \times  \sigma$. There is only a weak tendency
for redder nuclei in brighter objects, but again with a large uncertainty.

\subsection{Correlations among B-band structural parameters}

\begin{figure}
\centering
\includegraphics[width=9.0cm]{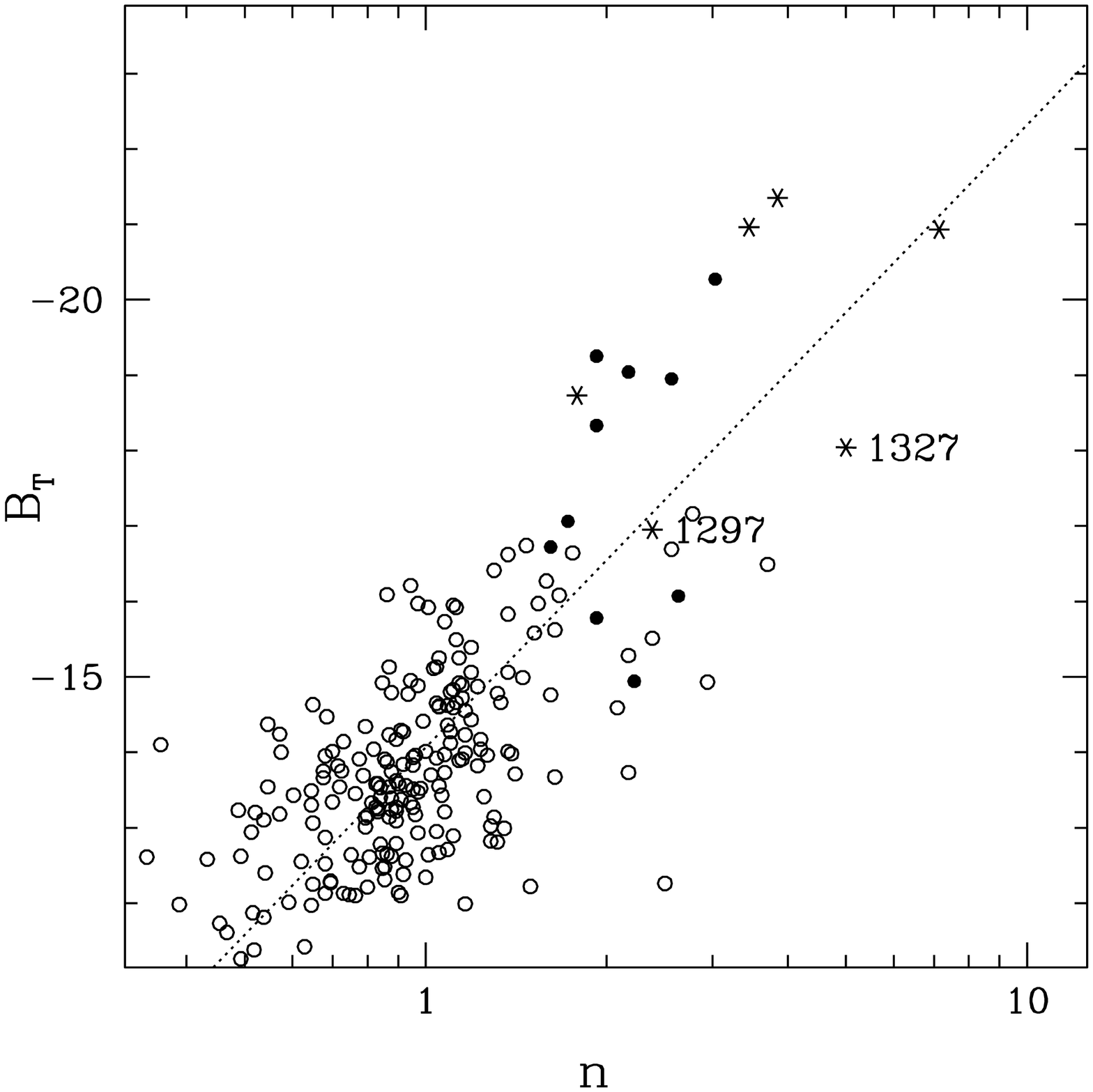}
\caption{The dependence of $n$ from $B_T$.
The dashed line gives the bisector linear regression.
}
\label{n_M}
\end{figure}

\begin{figure}
\centering
\includegraphics[width=7.5cm]{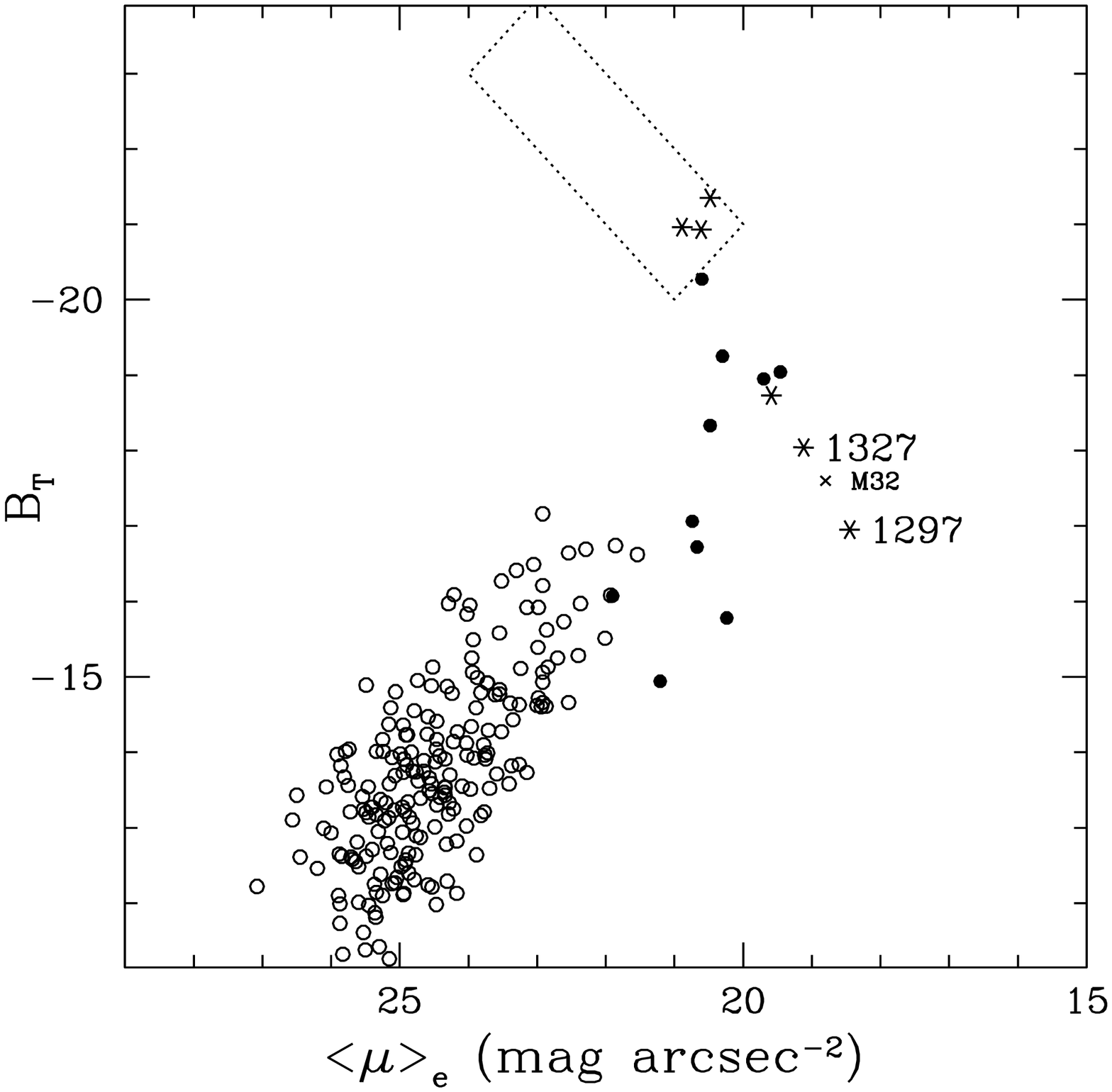}
\includegraphics[width=7.5cm]{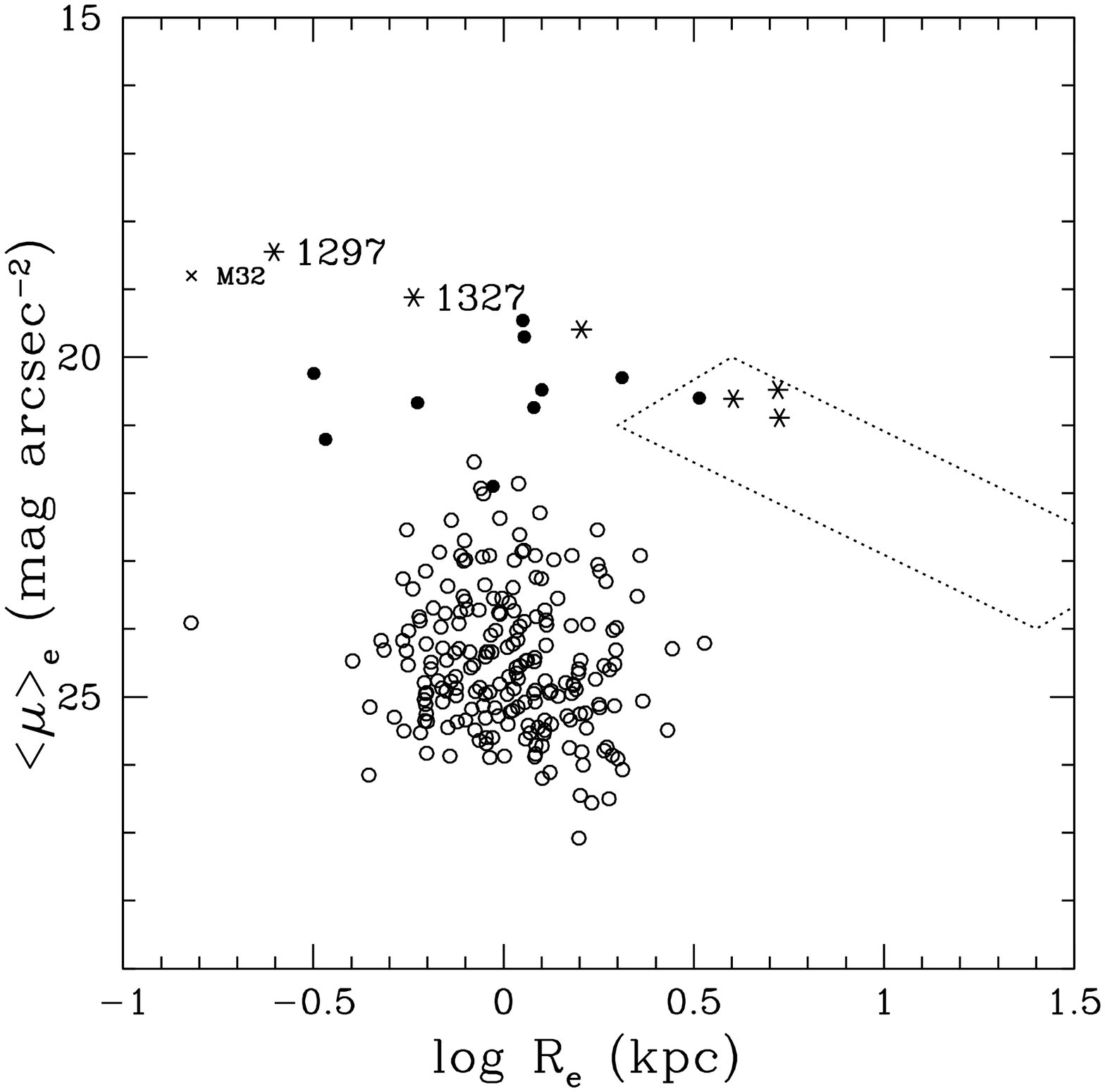}
\includegraphics[width=7.5cm]{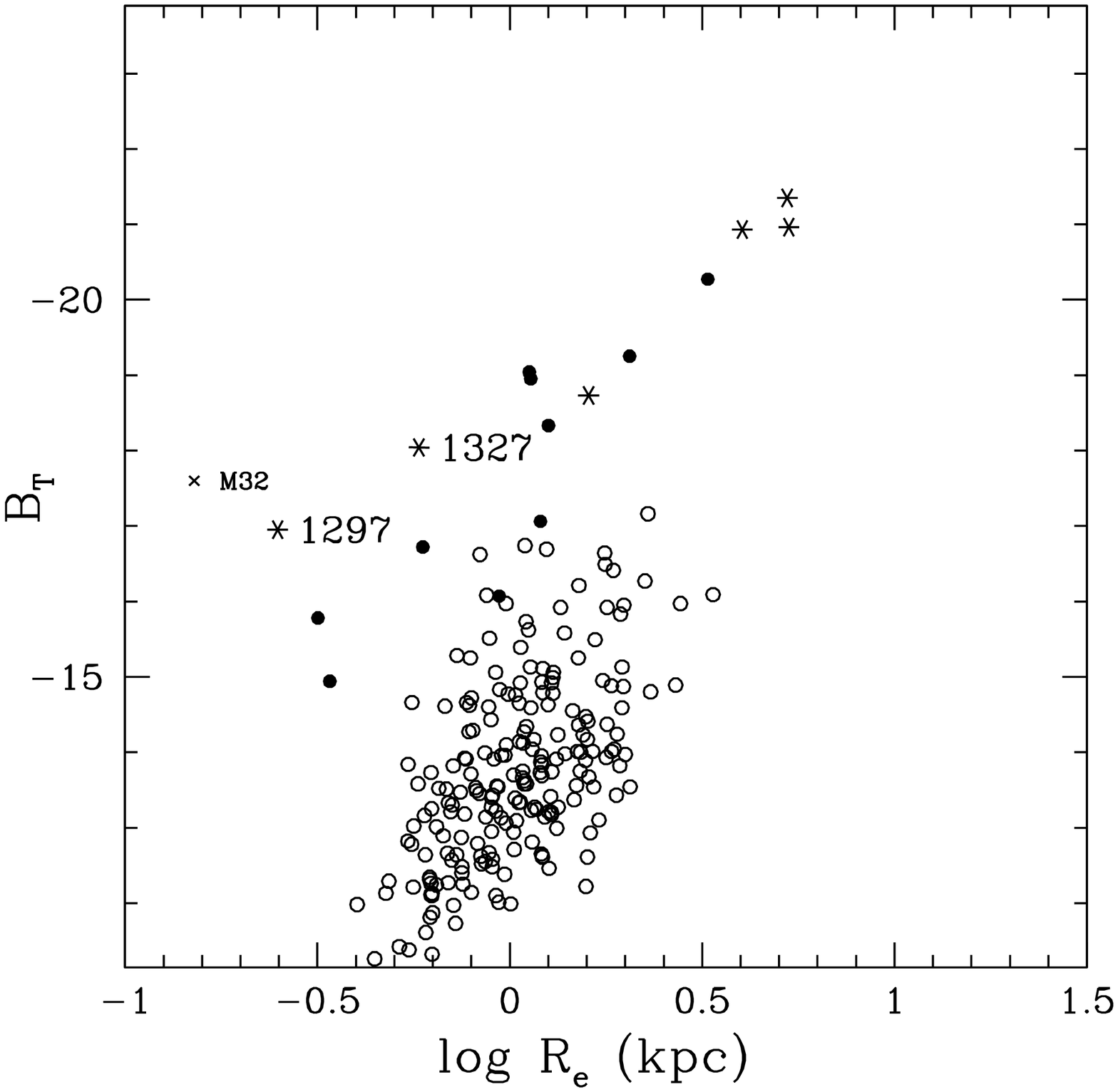}
\caption{Correlations among structural parameters of dE-E galaxies in the Virgo cluster.
$B_T$ vs. $<\mu>_e$ (top); $<\mu>_e$ vs. $log R_e$ (middle); 
$B_T$ vs. $log R_e$ (bottom).
The position of M32 (scaled to the distance of Virgo
and corrected for B-V=1) as given by Faber et al. (1997) is marked with 
a cross. The dotted rectangle represents the loci of luminous "core" galaxies adapted from GG03. 
}
\label{Guzman1}
\end{figure}

\begin{figure}
\centering
\includegraphics[width=7.5cm]{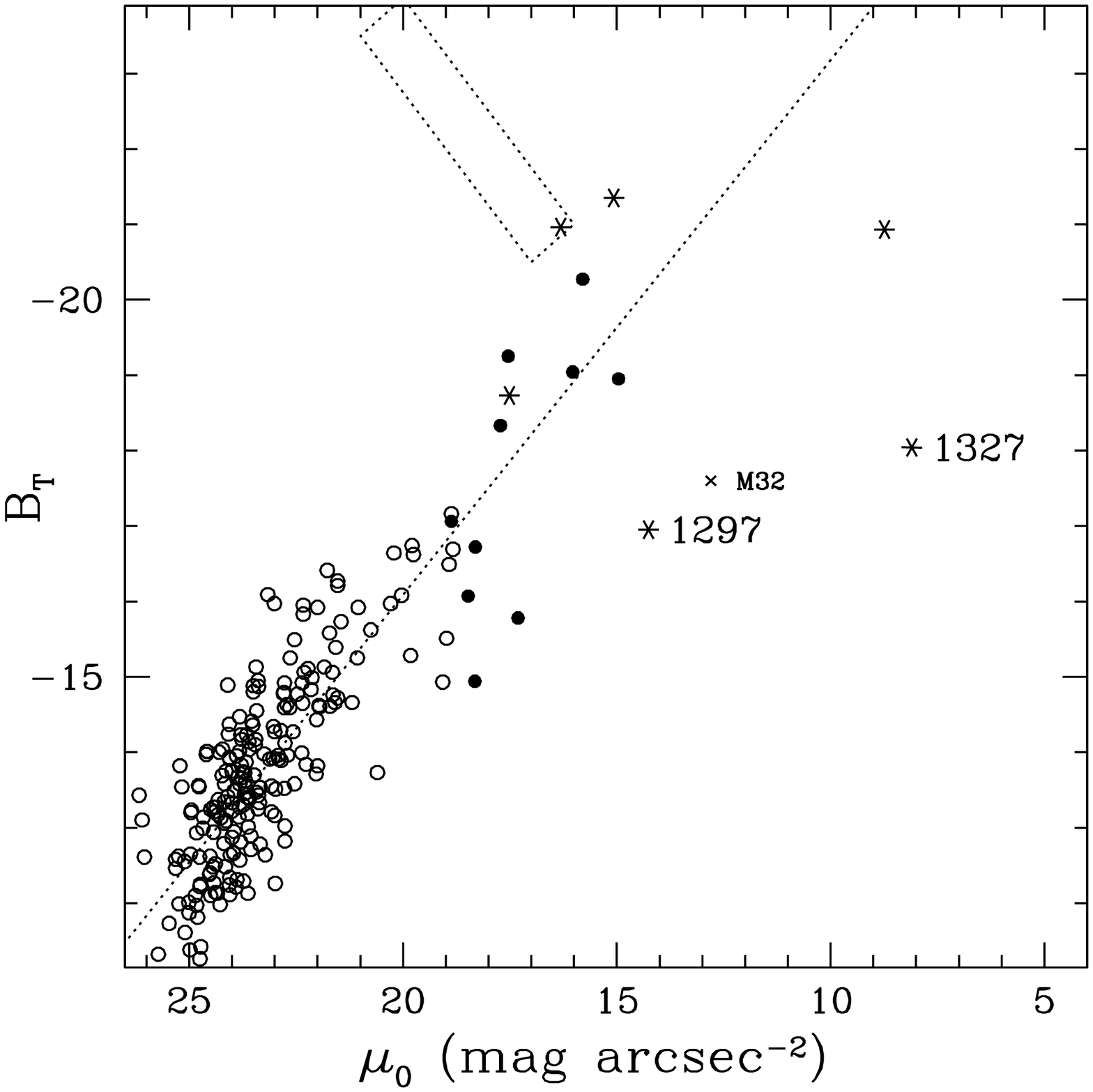}
\caption{Correlation between $B_T$ and $\mu_o$ for dE-E galaxies 
in the Virgo cluster. The position of M32 is marked with 
a cross. The dotted rectangle represents the loci of luminous "core" galaxies adapted from GG03. 
The dashed line gives the bisector linear regression. 
}
\label{Guzman2}
\end{figure}

\begin{figure}
\centering
\includegraphics[width=7.5cm]{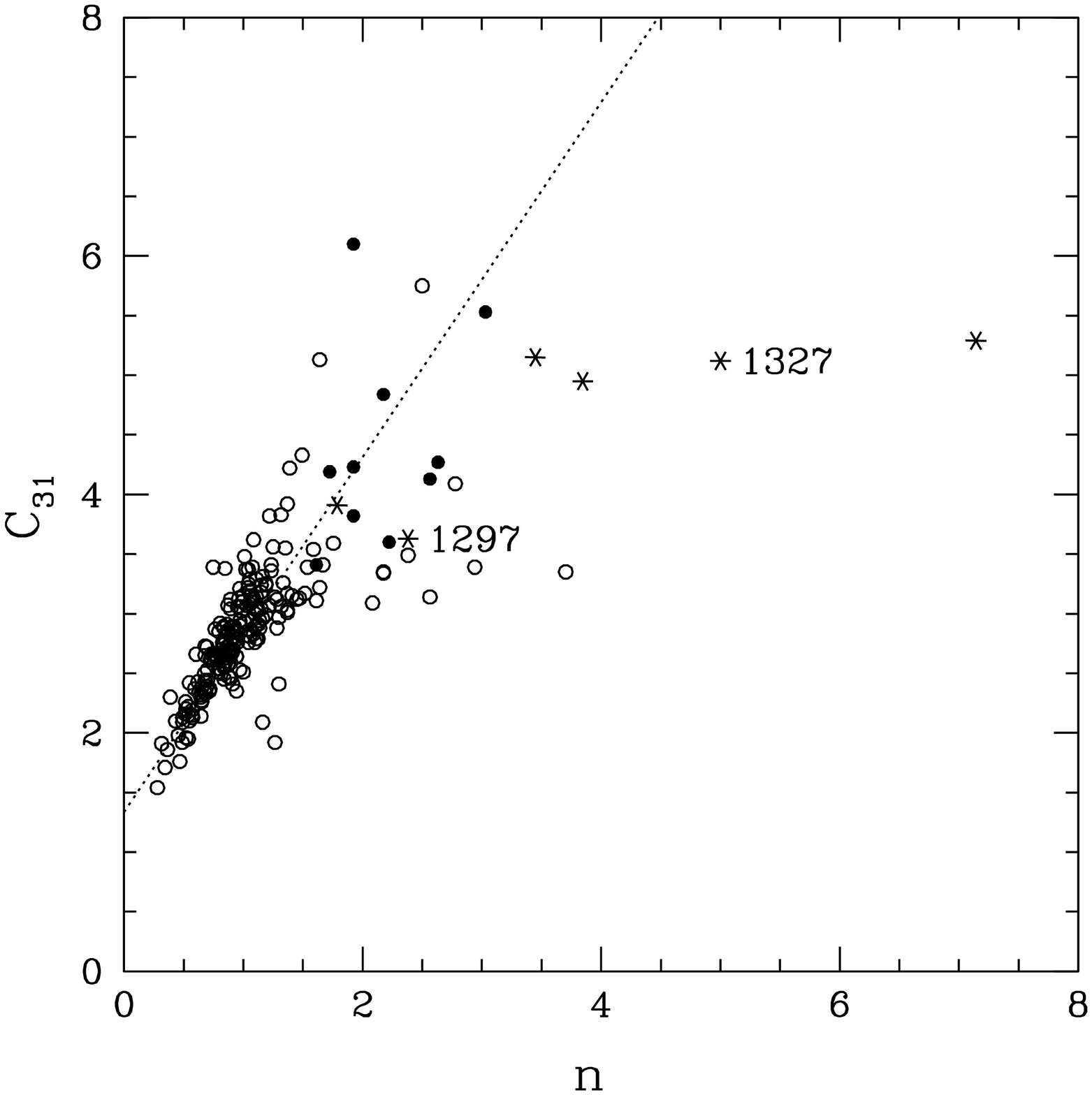}
\caption{Distribution of $C_{31}$ versus $n$.
The dashed line gives the bisector linear regression: $C_{31}= 1.49 \times n+1.33$
}
\label{c31}
\end{figure}

\begin{figure}
   \centering
   \includegraphics[width=18.0cm]{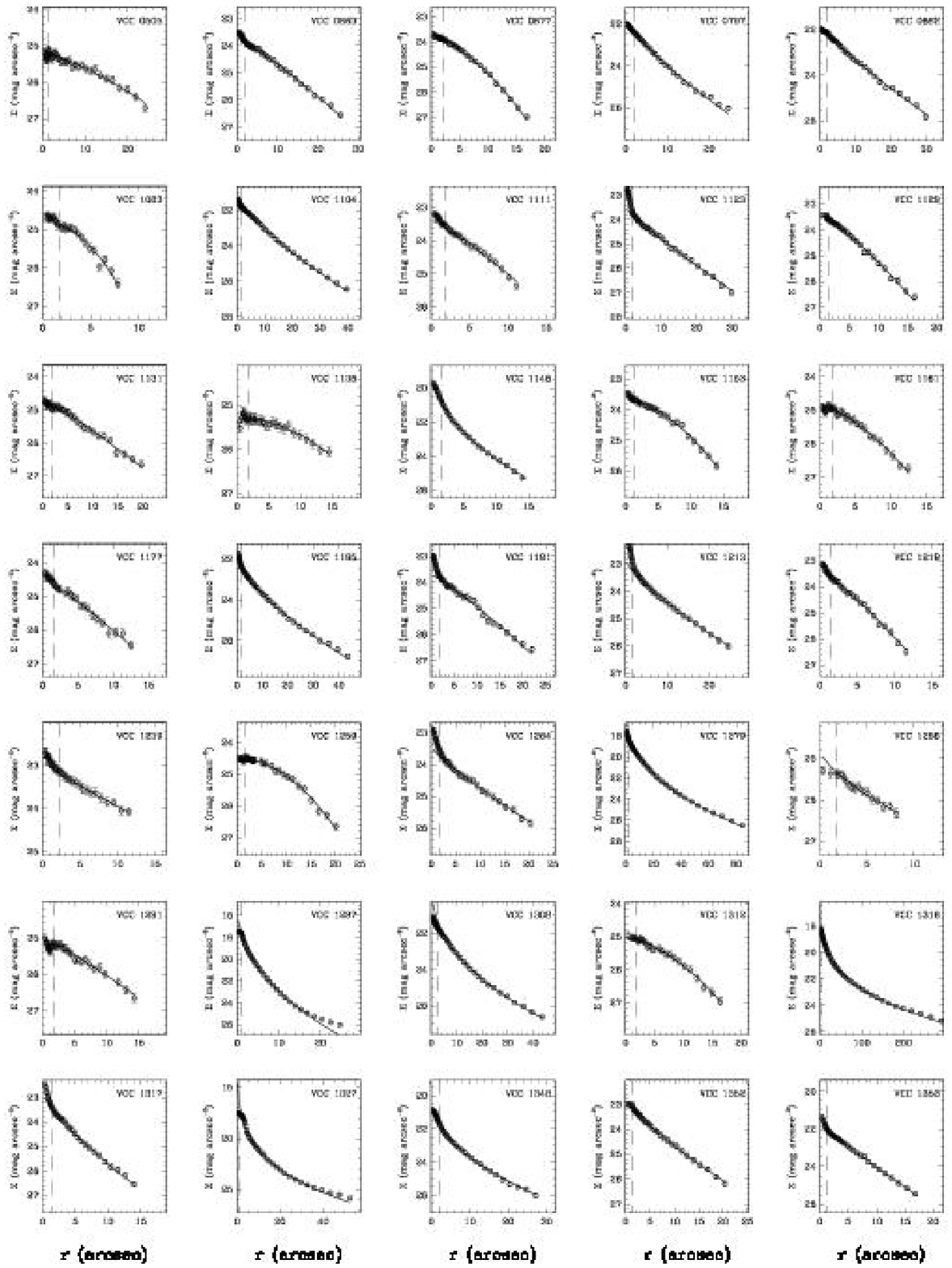}
   \caption{The 136 individual B-band profiles and their Sersic fit. The vertical dashed line marks the seeing. This is one page sample. 
   The entire figure is available at URL  http://goldmine.mib.infn.it/papers/WFS\_04/WFS\_04-frame.html}
   \label{profiles}
   \end{figure}

\begin{figure}
   \centering
   \includegraphics[width=18.0cm]{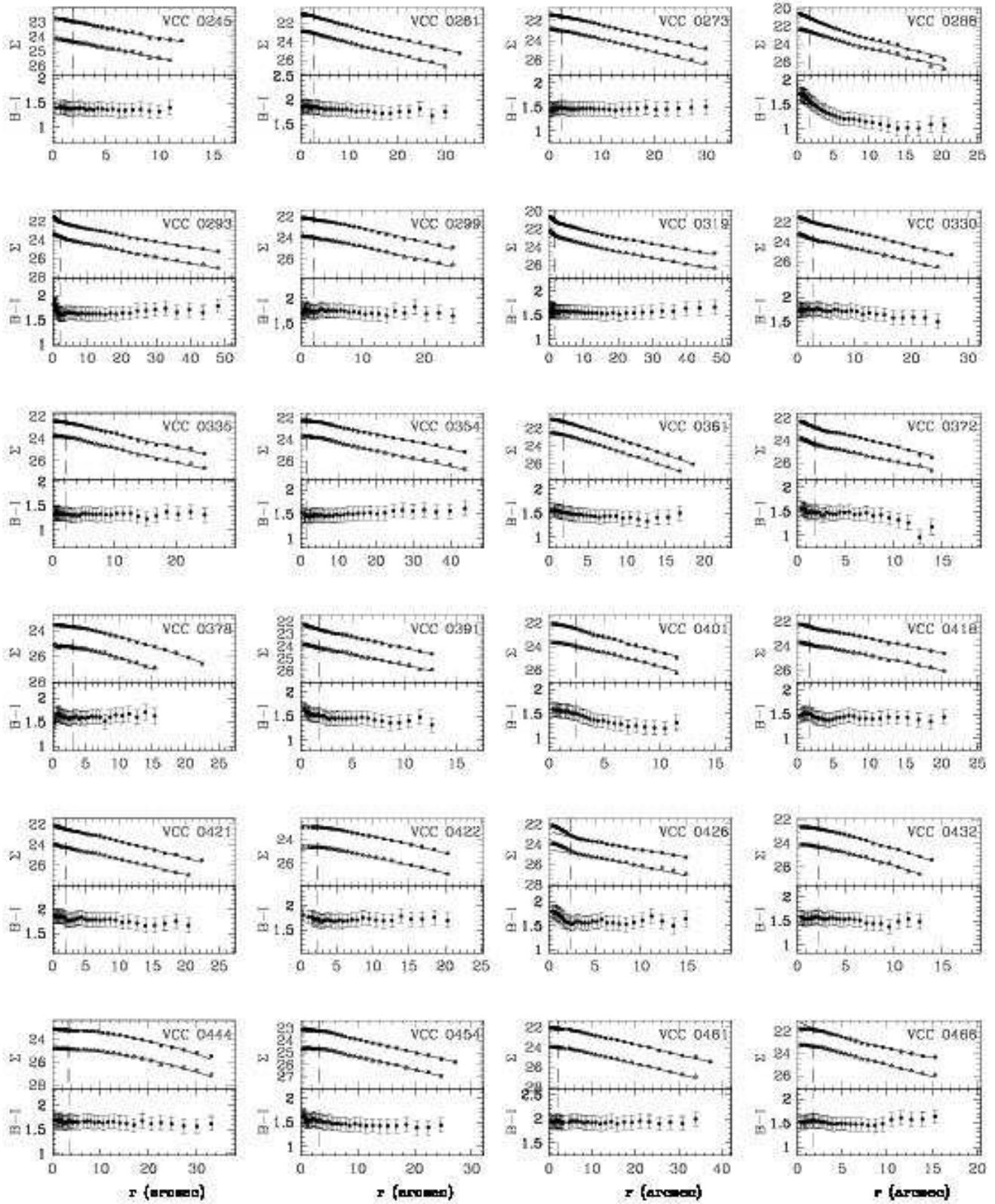}
   \caption{The profiles and their Sersic fit for 90 galaxies in the WFV with B and I band imaging and their B-I
   color profile.  I Images are convolved to the seeing of B images, marked by the vertical dashed line. This is one page sample. 
   The entire figure is available at URL http://goldmine.mib.infn.it/papers/WFS\_04/WFS\_04-frame.html}
   \label{profilesBI}
   \end{figure}

Guided by the work of GG03 who based their study   
on a sample of 18 dEs in Coma  
observed with the HST, combined with 232 elliptical galaxies taken from the literature, 
we analyze in this Section 
several correlations among structural parameters of dE-E galaxies in the Virgo cluster that are derived 
using the Sersic model.\\
We remark that the sample 
of Virgo elliptical galaxies we are using is 90\% complete at $m_p<19$. 
In other words the density of points in the various 
plots reflects the real frequency in the parameter space of Virgo galaxies.\\
We begin by remarking, in full agreement with GG03 and Caon et al. (1993), that 
in our sample there is a significant linear increase of the Sersic index $n$ with the system luminosity 
(see Fig.\ref{n_M}). Using the bisector linear regression we find:
\begin{equation}
log(n) = -0.12 \times  B_T +1.71 ~(R=-0.72)
\end{equation}
All dEs with $B_T>-17$ have Sersic index $n\lesssim 2$ while the few giant Es 
have $n$ as high as 7.\\
Secondly we show in Fig.\ref{Guzman1} (top) that the relation $B_T$ versus $<\mu>_e$, extensively studied 
by Binggeli \& Cameron (1991), Binggeli et al. (1984) and Ferguson \& Binggeli (1994) 
shows the existence of two separate regimes: dwarf elliptical galaxies
having brighter surface brightness with increasing luminosity, and giants showing the
reverse trend.
Furthermore the relation between the effective surface brightness and the 
radius (see Fig.\ref{Guzman1}, middle) is of inverse proportionality for giant ellipticals
(Kormendy \& Djorgowski 1989), i.e. smaller radii at brighter  
mean surface brightness, while dEs show a sparse relation (see also Capaccioli \& Caon 1991). Third we show (bottom)
that the scale $log R_e$ increases with $B_T$ more rapidly for giant Es than for dEs (see Binggeli et al. 1984).\\
In all panels of Fig.\ref{Guzman1} it appears that some dichotomy occurs
in passing from dEs to Es. However, as stressed for the first time by GG03, 
all bright galaxies showing deviant trends in this figure are
the "partially evacuated core" or ``core" galaxies that show a flat slope
in the inner $\lesssim$ 100 pc ($\lesssim 1.2$ arcsec at the distance of Virgo), 
as opposed to the "normal" galaxies with a central cusp 
(that Faber et al. 1997 define as "power-law" galaxies).   
It should be noted that the region of the Virgo cluster mapped with 
completeness in this work does not contain
many "core" galaxies, beside M84, M86 and M87 (the brightest objects in 
Fig.\ref{Guzman1}). Most "core" objects found in other clusters by Faber et 
al. (1997) are much more luminous ($-21<B_T<-24$) cD galaxies
with large outer envelopes that are absent in Virgo. These galaxies obey Sersic laws only in the
outer profiles and Graham et al. (2003b) have developed a formalism to model their profiles in their full extent.  
The rectangles in Fig.\ref{Guzman1} represent the loci occupied by them.
Notice that the M32-like objects (e.g. VCC 1297 and 1327) (both core galaxies) 
represent the low-luminosity continuation of the core regime.\\
The E-dE dichotomy no longer appears when $B_T$ is plotted as a function of $\mu_o$ in Fig.\ref{Guzman2}. 
GG03 conclude that "normal" dE-E galaxies have increasingly brighter central surface brightness with
increasing luminosity, until the onset of ``core" formation in elliptical galaxies at 
$B_T \sim -20.5$ mag. Together with 
GG03 we conclude that among "normal" E-dE galaxies there is no dichotomy in the structural parameters. 
Only the "core" galaxies seem structurally different, perhaps due to a different 
formation mechanism (see the discussion in Graham et al. 2003b).\\
Given the significant correlation between $B_T$ and $\log~n$, which clearly indicates a smooth,
continuous transition from the dE to the E regime,
and the (analytically\footnote{$C_{31}$ and $n$ are expected to have a perfect
correlation in the Sersic model. However in our definition 
$C_{31}$ is empirically determined from the data.} expected)
correlation between the light concentration index $C_{31}$ and the Sersic index $n$ (see Fig.\ref{c31}),
it is not surprising 
that $C_{31}$ increases linearly 
with the total luminosity $B_T$. This has been known since our Near-IR survey, as stressed by 
Scodeggio et al. (2002).\\
An additional element of continuity between
dE and E is the color--luminosity diagram of Fig.\ref{colorB-I} 
The gradual increase of the B-I index with increasing luminosity, 
resulting from the
metallicity--mass relation (Arimoto \& Yoshii 1987), encompasses the 
whole dE-E sequence, although with a larger scatter among dEs. \\

\section{Summary and conclusions}

We carried out a CCD survey of the North-East 
quadrant of the Virgo cluster containing $\sim$ 30\% of the galaxies in this cluster. 
We analyzed images for 226 and 90 galaxies respectively in the B- and I-band,  
representative of the properties of giant and dwarf elliptical galaxies in this cluster. \\
We fit the galaxies' radial light 
profiles with the Sersic $r^{1/n}$ model of light distribution obtaining
the structural parameters $n$, $<\mu>_e$, $r_e$, $B_T$, $C_{31}$ and $\mu_o$.
We find that the Sersic model provides an adequate representation of the
observed light profiles of dE and E galaxies spanning 9 magnitudes.\\
We confirm the result of Graham \& Guzman (2003) that the apparent dichotomy between 
E and dE galaxies in the luminosity--$<\mu>_e$ plane 
is due to the onset of ``core" formation in elliptical galaxies at 
$B_T \sim -20.5$ mag. The segregation among dE and E no longer appears when other structural parameters
are considered.\\
For 90 galaxies in the WFV we analyze the B-I color indices, both in the nuclear and in the outer regions.
Both indices are bluer toward fainter luminosities, however the scatter in these relations
increases significantly toward lower luminosities. 
Moreover we find that the outer color gradients do not show any significant correlation with the luminosity.

\begin{acknowledgements} 
We wish to thank Luca Cortese for the useful discussions.    
This work could not have been completed without the NASA/IPAC Extragalactic Database (NED) which 
is operated by the Jet Propulsion Laboratory, Caltech under contract with NASA.
This research has made use also of the GOLDMine Database (http://goldmine.mib.infn.it). 
\end{acknowledgements}

   \newpage
   \scriptsize
   \onecolumn
   \begin{longtable}{lccccccccrcccc}
   \caption{B-band parameters of the observed galaxies.}\\
   \hline
   \hline
   \noalign{\smallskip}
   VCC  & Type    &	  &     $m_p$   &  $n$  &  $\mu_o$   &  $B_T$   & $C_{31}$   & $<\mu>_e$       &  $r_e$  & Strip & Band & Notes & Ref \\
    (1)    &  (2) & (3)   &  (4)    &  (5)    & (6)    & (7)      & (8)     & (9)          & (10)      & (11)    & (12)    & (13)  & (14)  \\
   \noalign{\smallskip}
   \hline
   \noalign{\smallskip}
   \endfirsthead
   \caption{Continued}\\
   \hline
   \noalign{\smallskip}
   VCC  & Type    &	  &     $m_p$   &  $n$  &  $\mu_o$   &  $B_T$   & $C_{31}$   & $<\mu>_e$       &  $r_e$  & Strip & Band & Notes & Ref \\
    (1)    &  (2) & (3)   &  (4)    &  (5)    & (6)    & (7)      & (8)     & (9)          & (10)      & (11)    & (12)    & (13)  & (14)  \\
   \noalign{\smallskip}
   \hline
   \noalign{\smallskip}
   \endhead
   \hline
   \endfoot										               
 245  & dE & - & 18.40  & 0.89 & 24.14 & 18.06  & 2.46 & 25.22 & 12.57 &   V   & B,I &  	  &	       \\
 261  & dE & - & 16.00  & 1.11 & 22.65 & 16.56  & 3.29 & 23.89 & 13.69 &   V   & B,I &  	  &	       \\
 273  & dE & N & 16.55  & 0.88 & 22.78 & 16.36  & 2.58 & 23.82 & 14.71 &   V   & B,I &  	  &	       \\
 288  & dE & - & 17.65  & 1.22 & 22.00 & 17.33  & 3.07 & 23.37 &  8.61 &   V   & B,I &   1	  &	       \\
 293  & dE & N & 16.55  & 1.10 & 23.50 & 16.35  & 2.76 & 25.06 & 28.04 &   V   & B,I &  	  &	       \\
 299  & dE & - & 17.26  & 0.83 & 23.82 & 17.57  & 2.89 & 24.54 & 13.33 &   V   & B,I &  	  & BJ98       \\
 319  & dE & N & 16.15  & 1.37 & 22.33 & 15.32  & 3.01 & 24.02 & 23.41 &   V   & B,I &  	  & VPC        \\
 330  & dE & N & 16.65  & 0.89 & 23.45 & 16.98  & 2.75 & 24.46 & 13.95 &   V   & B,I &  	  & VPC        \\
 335  & dE & - & 17.75  & 1.02 & 23.48 & 17.45  & 3.37 & 24.27 & 12.34 &   V   & B,I &  	  &	       \\
 354  & dE & - & 16.55  & 1.04 & 23.44 & 16.02  & 3.37 & 24.52 & 23.65 &   V   & B,I &  	  & VPC,BJ98   \\
 361  & dE & - & 17.35  & 0.92 & 22.27 & 17.31  & 2.75 & 23.26 &  6.57 &   V   & B,I &  	  & VPC        \\
 372  & dE & N & 17.95  & 1.04 & 23.95 & 18.20  & 2.76 & 25.31 & 10.82 &   V   & B,I &  	  &	       \\
 378  & dE & - & 18.75  & 0.62 & 25.11 & 18.60  & 2.43 & 25.64 & 10.42 &   V   & B,I &  	  &	       \\
 391  & dE & - & 17.97  & 0.95 & 23.83 & 17.88  & 2.85 & 24.96 & 10.81 &   V   & B,I &  	  &	       \\
 401  & dE & - & 17.65  & 0.79 & 23.62 & 18.14  & 2.67 & 24.49 &  7.79 &   V   & B,I &  	  &	       \\
 418  & dE & - & 17.84  & 0.88 & 23.75 & 17.41  & 2.84 & 24.76 & 15.53 &   V   & B,I &  	  &	       \\
 421  & dE & - & 16.95  & 0.94 & 23.99 & 17.82  & 2.64 & 25.20 & 12.76 &   V   & B,I &  	  &	       \\
 422  & dE & - & 17.95  & 0.83 & 24.37 & 17.88  & 2.76 & 25.42 & 14.02 &   V   & B,I &  	  &	       \\
 426  & dE & N & 18.12  & 0.88 & 24.51 & 18.53  & 2.47 & 25.49 & 10.15 &   V   & B,I &  	  &	       \\
 432  & dE & - & 18.95  & 0.75 & 24.05 & 19.04  & 3.39 & 24.95 &  7.53 &   V   & B,I &  	  &	       \\
 444  & dE & - & 17.17  & 0.55 & 24.77 & 17.61  & 2.10 & 25.46 & 19.94 &   V   & B,I &  	  & BJ98       \\
 454  & dE & - & 17.55  & 0.89 & 24.43 & 17.88  & 3.12 & 25.40 & 16.11 &   V   & B,I &  	  &	       \\
 461  & dE & - & 16.45  & 0.87 & 23.79 & 16.92  & 2.68 & 24.89 & 18.74 &   V   & B,I &  	  & VPC        \\
 466  & dE & - & 18.07  & 0.95 & 22.98 & 17.64  & 2.76 & 23.97 &  8.27 &   V   & B,I &  	  & VPC        \\
 469  & dE & - & 18.50  & 0.80 & 23.00 & 17.99  & 2.67 & 23.82 &  7.25 &   V   & B,I &  	  &	       \\
 478  & dE & - & 18.58  & 0.80 & 23.90 & 18.94  & 2.54 & 24.53 &  6.77 &   V   & B,I &  	  &	       \\
 481  & dE & - & 18.30  & 0.34 & 26.05 & 18.54  & 1.71 & 26.45 & 19.20 &   V   & B,I &  	  &	       \\
 493  & dE & - & 18.95  & 0.63 & 24.73 & 19.73  & 2.34 & 25.30 &  6.23 &   V   & B,I &  	  &	       \\
 494  & dE & - & 16.59  & 0.68 & 23.82 & 16.68  & 2.33 & 24.59 & 19.03 &   V   & B,I &  	  & VPC,BJ98   \\
 505  & dE & N & 17.75  & 0.72 & 25.17 & 17.61  & 2.37 & 26.07 & 24.79 &   V   & B   &  	  &	       \\
 510  & dE & N & 15.07  & 1.11 & 22.33 & 15.20  & 3.14 & 23.98 & 23.90 &   V   & B,I &  	  & BJ98       \\
 515  & dE & - & 18.15  & 0.70 & 24.58 & 17.14  & 2.44 & 25.24 & 19.81 &   V   & B,I &  	  &	       \\
 521  & dE & - & 18.45  & 0.97 & 24.83 & 18.22  & 3.21 & 26.00 & 19.56 &   V   & B,I &  	  &	       \\
 535  & dE & - & 16.95  & 1.22 & 23.38 & 16.28  & 3.82 & 24.31 & 23.81 &   V   & B,I &  	  &	       \\
 536  & dE & - & 18.95  & 0.31 & 25.72 & 19.83  & 1.91 & 25.83 &  7.58 &   V   & B,I &  	  &	       \\
 539  & dE & N & 16.75  & 0.73 & 23.60 & 17.01  & 2.62 & 24.22 & 12.80 &   V   & B,I &  	  & VPC        \\
 543  & dE & - & 14.33  & 1.75 & 20.21 & 14.51  & 3.59 & 22.54 & 21.30 &   V   & B,I &  	  & ACS,BJ98   \\
 545  & dE & N & 15.19  & 1.64 & 20.75 & 15.53  & 3.22 & 22.86 & 13.49 &   V   & B,I &   3	  &	       \\
 547  & dE & - & 18.75  & 0.52 & 25.00 & 19.28  & 2.26 & 25.36 &  7.61 &   V   & B,I &  	  &	       \\
 551  & dE & - & 16.30  & 0.55 & 24.06 & 16.78  & 2.42 & 25.16 & 21.61 &   V   & B,I &  	  &	       \\
 554  & dE & N & 17.06  & 0.85 & 24.05 & 17.24  & 2.60 & 24.94 & 15.96 &   V   & B,I &  	  & BJ98       \\
 563  & dE & - & 16.25  & 0.36 & 23.46 & 17.05  & 1.86 & 23.78 & 11.83 &   V   & B,I &  	  &	       \\
 592  & dE & N & 16.55  & 1.05 & 21.95 & 16.55  & 3.29 & 22.94 & 10.64 &   V   & B,I &  	  & VPC        \\
 594  & dE & - & 17.08  & 0.68 & 24.14 & 17.40  & 2.50 & 24.81 & 18.41 &   V   & B,I &  	  & BJ98       \\
 600  & dE & - & 17.95  & 0.91 & 24.86 & 19.05  & 2.41 & 25.89 & 11.12 &   V   & B,I &  	  &	       \\
 608  & dE & N & 14.86  & 0.94 & 21.53 & 14.94  & 2.64 & 22.92 & 18.27 &   V   & B,I &  	  & BJ98       \\
 614  & dE & - & 18.55  & 0.39 & 24.28 & 19.17  & 2.30 & 24.47 &  4.84 &   V   & B,I &  	  &	       \\
 619  & dE & - & 18.15  & 0.65 & 24.17 & 18.09  & 2.26 & 24.81 & 11.79 &   V   & B,I &  	  &	       \\
 625  & dE & - & 18.15  & 0.88 & 23.59 & 17.76  & 2.85 & 24.70 & 12.47 &   V   & B,I &  	  &	       \\
 634  & dE & N & 14.08  & 2.78 & 18.87 & 13.99  & 4.09 & 22.92 & 27.61 &   V   & B,I &  	  &	       \\
 635  & dE & - & 18.95  & 1.09 & 23.56 & 18.44  & 3.62 & 25.40 & 12.40 &   V   & B,I &  	  &	       \\
 643  & dE & - & 18.45  & 0.76 & 24.50 & 19.05  & 2.87 & 25.25 &  7.57 &   V   & B,I &  	  &	       \\
 646  & dE & - & 18.75  & 0.76 & 23.70 & 17.70  & 2.65 & 24.53 & 10.07 &   V   & B,I &  	  &	       \\
 663  & dE & - & 18.15  & 0.86 & 23.67 & 17.28  & 2.91 & 24.48 & 14.53 &   V   & B   &  	  &	       \\
 668  & dE & - & 16.45  & 1.08 & 23.70 & 17.42  & 3.39 & 24.95 & 14.49 &   V   & B,I &  	  & VPC        \\
 674  & dE & N & 17.95  & 1.00 & 24.59 & 17.14  & 3.15 & 25.79 & 22.20 &   V   & B,I &  	  &	       \\
 677  & dE & - & 18.15  & 0.65 & 23.74 & 17.85  & 2.32 & 24.46 &  8.56 &   V   & B   &  	  &	       \\
 684  & dE & N & 15.98  & 1.11 & 22.15 & 16.32  & 2.79 & 23.55 & 11.37 &   V   & B,I &  	  & VPC,BJ98   \\
 696  & dE & - & 18.25  & 0.84 & 23.64 & 17.75  & 2.55 & 24.41 & 10.83 &   V   & B,I &  	  &	       \\
 708  & dE & N & 18.95  & 1.00 & 24.05 & 18.81  & 2.51 & 25.04 &  7.48 &   V   & B,I &  	  &	       \\
 714  & dE & - & 18.25  & 0.68 & 24.00 & 18.28  & 2.44 & 24.70 &  9.03 &   V   & B,I &  	  &	       \\
 725  & dE & N & 15.95  & 0.94 & 23.39 & 16.20  & 2.35 & 24.74 & 21.08 &   V   & B,I &  	  &	       \\
 726  & dE & N & 18.65  & 1.16 & 25.24 & 19.16  & 2.09 & 25.87 & 12.14 &   V   & B,I &  	  &	       \\
 748  & dE & - & 17.25  & 0.89 & 23.68 & 17.53  & 3.04 & 24.73 & 13.20 &   V   & B,I &  	  & VPC        \\
 754  & dE & - & 18.95  & 0.54 & 24.80 & 19.34  & 1.95 & 25.35 &  7.49 &   V   & B,I &  	  &	       \\
 757  & dE & - & 18.55  & 0.93 & 23.82 & 18.58  & 2.89 & 24.91 &  8.53 &   V   & B,I &  	  &	       \\
 761  & dE & - & 17.21  & 0.95 & 24.06 & 17.22  & 3.06 & 25.11 & 21.51 &   V   & B,I &  	  & BJ98       \\
 763 M84 &  E & - & 10.20  & 7.14 &  8.74 & 10.22  & 5.29 & 20.61 & 48.55 &   V   & B,I &  3,6,4	  & ACS*       \\
 765  & dE & N & 16.43  & 1.12 & 21.19 & 16.49  & 2.79 & 22.54 &  6.72 &   V   & B,I &  	  & VPC,BJ98   \\
 777  & dE & - & 17.95  & 0.88 & 24.50 & 17.91  & 2.65 & 25.53 & 14.17 &   V   & B,I &  	  &	       \\
 779  & dE & N & 17.62  & 1.25 & 24.10 & 17.74  & 3.56 & 25.54 & 15.42 &   V   & B,I &  	  & BJ98       \\
 780  & dE & - & 18.27  & 2.50 & 22.99 & 18.89  & 5.75 & 25.11 &  7.53 &   V   & B,I &  	  &	       \\
 789  & dE & - & 18.95  & 0.81 & 24.76 & 18.54  & 2.92 & 25.71 & 14.69 &   V   & B,I &  	  &	       \\
 790  & dE & N & 16.34  & 1.14 & 21.07 & 15.90  & 2.88 & 22.70 &  9.53 &   V   & B,I &  	  & VPC        \\
 795  & dE & N & 17.95  & 1.05 & 23.08 & 17.60  & 3.19 & 24.09 & 11.17 &   V   & B,I &  	  & VPC        \\
 797  & dE & N & 16.94  & 1.33 & 21.59 & 16.49  & 3.26 & 22.92 &  9.32 &   V   & B   &  	  &	       \\
 808  & dE & N & 17.65  & 0.99 & 23.54 & 16.74  & 3.10 & 24.46 & 19.25 &   V   & B,I &  	  & VPC        \\
 810  & dE & N & 16.91  & 0.91 & 22.86 & 16.86  & 2.68 & 23.71 &  9.67 &   V   & B,I &  	  & B03,VPC,BJ98\\
 812  & dE & N & 16.98  & 1.10 & 23.00 & 16.88  & 2.89 & 24.16 & 13.15 &   V   & B,I &  	  & VPC        \\
 813  & dE & N & 18.15  & 0.93 & 24.78 & 17.59  & 2.84 & 25.75 & 18.01 &   V   & B,I &  	  &	       \\
 815  & dE & N & 16.06  & 1.45 & 22.13 & 16.16  & 3.12 & 23.87 & 15.61 &   V   & B,I &  	  & B03.VPC,BJ98\\
 816  & dE & N & 15.25  & 0.97 & 23.01 & 15.18  & 2.53 & 24.29 & 33.51 &   V   & B,I &  	  &	       \\
 817  & dE & - & 14.93  & 3.70 & 18.92 & 14.66  & 3.35 & 23.05 & 21.35 &   V   & B,I &  	  &	       \\
 824  & dE & N & 17.95  & 2.08 & 22.77 & 16.56  & 3.09 & 25.13 & 23.64 &   V   & B,I &  	  &	       \\
 829  & dE & - & 18.45  & 0.90 & 24.39 & 19.01  & 2.89 & 25.34 &  9.59 &   V   & B,I &  	  &	       \\
 833  & dE & N & 17.42  & 1.15 & 23.12 & 17.24  & 3.04 & 24.34 & 10.97 &   V   & B,I &  	  & VPC        \\
 844  & dE & - & 18.85  & 0.52 & 24.98 & 19.77  & 2.20 & 25.50 &  6.61 &   V   & B,I &  	  &	       \\
 845  & dE & - & 18.95  & 0.92 & 24.52 & 18.77  & 2.89 & 25.28 & 11.71 &   V   & B,I &  	  &	       \\
 846  & dE & N & 16.16  & 1.61 & 21.64 & 16.39  & 3.11 & 23.61 & 12.50 &   V   & B,I &  	  & B03,VPC    \\
 861  & dE & - & 17.85  & 1.06 & 23.38 & 17.72  & 3.15 & 24.34 & 10.90 &   V   & B,I &  	  &	       \\
 872  & dE & N & 16.93  & 1.04 & 23.04 & 17.23  & 3.25 & 23.92 &  9.21 &   V   & B,I &  	  & BJ98       \\
 877  & dE & N & 17.55  & 0.83 & 23.39 & 17.90  & 2.58 & 24.22 &  7.56 &   V   & B,I &  	  &	       \\
 878  & dE & - & 17.25  & 0.57 & 24.30 & 17.15  & 2.13 & 24.83 & 18.44 &   V   & B,I &  	  &	       \\
 881 M86 &  E & - & 10.01  & 3.45 & 16.31 & 10.19  & 5.15 & 20.89 & 64.16 &   V   & B,I &  3,4	  & ACS*       \\
 882  & dE & N & 16.63  & 1.19 & 21.57 & 15.76  & 2.99 & 22.99 & 12.88 &   V   & B   &  	  &	       \\
1083  & dE & - & 19.00  & 0.49 & 24.75 & 19.89  & 2.13 & 25.15 &  5.38 &   H   & B   &  	  &	       \\
1104  & dE & N & 15.14  & 1.08 & 21.45 & 15.42  & 3.12 & 22.61 & 13.30 &   H   & B   &  3	  & B03,VPC,BJ98\\
1111  & dE & N & 17.65  & 0.84 & 23.34 & 17.62  & 2.45 & 24.34 &  9.85 &   H   & B   &  	  &	       \\
1123  & dE & N & 16.60  & 1.09 & 23.51 & 16.79  & 2.83 & 24.95 & 18.22 &   H   & B   &  	  & VPC        \\
1129  & dE & - & 17.64  & 0.81 & 23.36 & 17.82  & 2.50 & 24.28 &  8.36 &   H   & B   &  	  & B03        \\
1131  & dE & - & 18.05  & 0.87 & 24.67 & 18.01  & 3.07 & 25.45 & 14.89 &   H   & B   &  	  &	       \\
1136  & dE & N & 17.97  & 0.49 & 25.25 & 18.53  & 2.09 & 25.84 & 14.65 &   H   & B   &  	  &	       \\
1148  &  E & - & 15.90  & 2.22 & 18.32 & 16.21  & 3.60 & 21.21 &  4.11 &   H   & B   &  	  & VPC        \\
1153  & dE & - & 17.75  & 0.57 & 23.64 & 17.97  & 2.19 & 24.29 &  9.22 &   H   & B   &  	  &	       \\
1161  & dE & - & 18.98  & 0.65 & 24.83 & 19.18  & 2.32 & 25.45 &  8.62 &   H   & B   &  	  &	       \\
1177  & dE & - & 18.65  & 0.85 & 24.44 & 18.67  & 2.77 & 25.60 & 10.87 &   H   & B   &  	  &	       \\
1185  & dE & N & 15.62  & 1.52 & 21.72 & 15.57  & 3.17 & 23.55 & 16.76 &   H   & B   &  	  & VPC,ACS,BJ98\\
1191  & dE & N & 17.37  & 0.87 & 23.68 & 17.61  & 2.86 & 24.34 & 11.25 &   H   & B   &  	  &	       \\
1213  & dE & N & 16.37  & 1.32 & 22.80 & 16.37  & 3.06 & 24.24 & 15.61 &   H   & B   &  	  & VPC,BJ98   \\
1219  & dE & N & 18.16  & 0.84 & 23.34 & 18.37  & 2.88 & 24.32 &  6.71 &   H   & B   &  	  &	       \\
1239  & dE & N & 17.75  & 1.14 & 22.85 & 17.26  &19.21 & 24.65 & 19.00 &   H   & B   &  	  &	       \\
1259  & dE & - & 18.19  & 0.51 & 24.43 & 18.21  & 2.16 & 24.96 & 12.36 &   H   & B   &  	  &	       \\
1264  & dE & N & 17.26  & 1.16 & 23.42 & 16.60  & 2.96 & 24.79 & 17.60 &   H   & B   &  	  & BJ98       \\
1279  &  E & - & 12.08  & 2.56 & 14.96 & 12.20  & 4.13 & 19.70 & 13.68 &   H   & B   &   3,5	  & F97,ACS    \\
1286  & dE & - & 18.80  & 1.49 & 24.75 & 18.93  & 4.33 & 27.08 & 19.05 &   H   & B   &  	  &	       \\
1291  & dE & - & 18.77  & 0.86 & 24.97 & 18.50  & 2.63 & 25.88 & 14.60 &   H   & B   &  	  &	       \\
1297  &  E & - & 14.26  & 2.38 & 14.26 & 14.20  & 3.63 & 18.45 &  3.01 &   H   & B   &   1,2,4    & F97,ACS    \\
1308  & dE & N & 15.56  & 2.38 & 18.98 & 15.64  & 3.49 & 22.01 & 10.69 &   H   & B   &   3	  & VPC,BJ98   \\
1312  & dE & - & 18.65  & 0.59 & 25.01 & 19.14  & 2.37 & 25.60 & 11.29 &   H   & B   &  	  &	       \\
1316 M87  &  E & - &  9.51  & 3.85 & 15.07 &  9.80  & 4.95 & 20.48 & 63.53 &   H   & B   &   4	  & F97.ACS    \\
1317  & dE & N & 17.93  & 1.28 & 22.76 & 18.13  & 3.12 & 24.03 &  6.79 &   H   & B   &  	  &	       \\
1327  &  E & - & 13.20  & 5.00 &  8.10 & 13.11  & 5.12 & 19.12 &  7.01 &   H   & B   &   1,2,4    & ACS*       \\
1348  & dE & - & 15.79  & 2.17 & 19.82 & 15.87  & 3.34 & 22.40 &  8.80 &   H   & B   &  	  & VPC,BJ98   \\
1352  & dE & - & 17.15  & 1.27 & 22.70 & 17.19  & 3.14 & 24.02 & 11.51 &   H   & B   &  	  & VPC        \\
1353  & dE & N & 16.56  & 1.05 & 21.71 & 16.54  & 3.08 & 22.87 &  8.20 &   H   & B   &  	  & VPC,BJ98   \\
1363  & dE & N & 18.93  & 0.69 & 23.73 & 18.86  & 2.72 & 24.31 &  5.86 &   H   & B   &  	  &	       \\
1366  & dE & N & 17.55  & 1.23 & 23.77 & 16.98  & 3.41 & 25.25 & 19.19 &   H   & B   &  	  &	       \\
1369  & dE & N & 17.23  & 1.10 & 22.76 & 17.03  & 3.00 & 24.03 & 13.07 &   H   & B   &  	  & VPC        \\
1370  & dE & - & 17.35  & 0.65 & 23.95 & 17.66  & 2.14 & 24.57 &  9.89 &   H   & B   &  	  &	       \\
1381  & dE & - & 18.95  & 0.65 & 24.74 & 18.90  & 2.27 & 25.37 &  9.12 &   H   & B   &  	  &	       \\
1386  & dE & N & 14.25  & 1.59 & 21.53 & 14.88  & 3.54 & 23.52 & 27.11 &   H   & B   &  	  & VPC        \\
1389  & dE & N & 15.85  & 1.19 & 21.65 & 16.09  & 3.25 & 22.92 & 11.09 &   H   & B   &  	  & VPC,BJ98   \\
1396  & dE & N & 17.16  & 0.83 & 24.17 & 17.57  & 2.57 & 25.15 & 13.17 &   H   & B   &  	  & BJ98       \\
1399  & dE & N & 16.44  & 0.96 & 22.92 & 17.19  & 3.13 & 23.76 & 11.73 &   H   & B   &  	  & BJ98       \\
1402  & dE & N & 17.95  & 0.89 & 24.19 & 18.36  & 2.65 & 25.18 &  9.95 &   H   & B   &  	  &	       \\
1405  & dE & - & 18.95  & 1.27 & 25.70 & 20.76  & 1.92 & 26.15 &  5.34 &   H   & B   &  	  &	       \\
1407  & dE & N & 15.43  & 1.54 & 20.29 & 15.18  & 3.39 & 22.37 & 11.81 &   H   & B   &   3	  & VPC,ACS,BJ98\\
1414  & dE & - & 16.95  & 2.17 & 20.60 & 17.42  & 3.35 & 23.15 &  7.53 &   H   & B   &   3	  &	       \\
1418  & dE & - & 17.37  & 0.68 & 23.77 & 17.49  & 2.37 & 24.57 & 13.03 &   H   & B   &  	  &	       \\
1420  & dE & N & 16.34  & 1.09 & 21.96 & 16.53  & 3.13 & 23.00 &  9.48 &   H   & B   &  	  & VPC,BJ98   \\
1431  & dE & N & 14.43  & 1.37 & 19.76 & 14.53  & 3.17 & 21.54 & 10.10 &   H   & B   &  	  & ACS        \\
1438  & dE & - & 17.82  & 0.54 & 26.10 & 18.05  & 2.22 & 26.56 & 20.60 &   H   & B   &  	  &	       \\
1445  & dE & - & 18.37  & 1.28 & 22.76 & 18.33  & 2.88 & 24.17 &  6.56 &   H   & B   &  	  &	       \\
1454  & dE & N & 18.64  & 0.43 & 25.32 & 18.57  & 2.10 & 25.69 & 10.88 &   H   & B   &  	  &	       \\
1463  & dE & - & 18.45  & 1.01 & 23.22 & 18.51  & 2.93 & 23.88 &  7.29 &   H   & B   &  	  &	       \\
1464  & dE & - & 17.67  & 0.70 & 24.18 & 17.81  & 2.53 & 24.88 & 12.84 &   H   & B   &  	  &	       \\
1489  & dE & - & 15.84  & 1.03 & 22.22 & 16.04  & 3.07 & 23.24 & 14.72 &   H   & B   &  	  & VPC,ACS    \\
1517  & dE & N & 17.25  & 1.30 & 23.84 & 18.01  & 2.41 & 24.86 & 10.46 &   H   & B   &  	  &	       \\
1521  &  E & - & 14.14  & 1.72 & 18.87 & 14.09  & 4.19 & 20.74 & 14.51 &   H   & B   &  	  &	       \\
1523  & dE & N & 17.59  & 0.97 & 23.43 & 17.68  & 2.94 & 24.35 &  8.98 &   H   & B   &  	  & BJ98       \\
1536  & dE & N & 18.35  & 0.85 & 25.32 & 18.69  & 3.38 & 26.20 & 15.28 &   H   & B   &  	  &	       \\
1539  & dE & N & 15.61  & 1.37 & 22.31 & 16.09  & 3.03 & 23.94 & 15.67 &   H   & B   &  	  & VPC,ACS,BJ98\\
1545  &  E & - & 14.88  & 2.63 & 18.48 & 15.08  & 4.27 & 21.90 & 11.34 &   H   & B   &  	  & ACS        \\
1548  & dE & - & 18.54  & 0.68 & 23.63 & 19.02  & 2.34 & 24.17 &  5.75 &   H   & B   &  	  &	       \\
1549  & dE & N & 14.56  & 1.67 & 20.03 & 15.07  & 3.41 & 21.93 & 10.51 &   H   & B   &  	  & VPC        \\
1563  & dE & N & 16.06  & 0.97 & 23.50 & 16.27  & 2.90 & 24.54 & 22.19 &   H   & B   &  	  & VPC,BJ98   \\
1565  & dE & N & 16.86  & 1.23 & 24.22 & 17.11  & 3.36 & 25.74 & 22.57 &   H   & B   &  	  & VPC        \\
1594  & dE & - & 18.68  & 0.28 & 24.06 & 18.91  & 1.54 & 24.59 &  7.78 &   H   & B   &  	  &	       \\
1595  & dE & - & 18.33  & 0.85 & 23.96 & 18.49  & 2.68 & 24.87 &  8.32 &   H   & B   &  	  &	       \\
1599  & dE & - & 17.25  & 0.52 & 24.96 & 17.95  & 1.96 & 25.49 & 15.50 &   H   & B   &  	  &	       \\
1606  & dE & N & 17.45  & 1.39 & 23.25 & 17.17  & 4.22 & 24.99 & 16.81 &   H   & B   &  	  &	       \\
1609  & dE & N & 17.32  & 1.37 & 23.83 & 17.14  & 3.92 & 25.34 & 18.09 &   H   & B   &  	  &	       \\
1613  & dE & - & 18.45  & 0.78 & 24.15 & 18.67  & 2.63 & 24.98 &  9.07 &   H   & B   &  	  &	       \\
1619  &  E & - & 12.44  & 1.79 & 17.51 & 12.42  & 3.91 & 19.59 & 19.36 &   H   & B   &  4	  & ACS*       \\
1621  & dE & - & 18.32  & 0.73 & 24.34 & 19.02  & 2.66 & 24.94 &  7.58 &   H   & B   &  	  &	       \\
1626  & dE & - & 18.95  & 0.85 & 23.88 & 18.84  & 2.60 & 24.79 &  7.48 &   H   & B   &  	  &	       \\
1627  &  E & - & 15.11  & 1.92 & 17.31 & 15.37  & 3.82 & 20.24 &  3.83 &   H   & B   &  	  & ACS        \\
1630  &  E & - & 12.84  & 1.92 & 17.72 & 12.82  & 4.23 & 20.48 & 15.21 &   H   & B   &   5	  & F97,ACS    \\
1637  & dE & N & 18.45  & 1.11 & 23.56 & 18.26  & 3.03 & 24.76 &  8.11 &   H   & B   &  	  &	       \\
1642  & dE & N & 17.75  & 1.64 & 23.86 & 17.48  & 5.13 & 25.81 & 19.36 &   H   & B   &  	  &	       \\
1647  & dE & - & 15.95  & 0.85 & 22.77 & 16.23  & 2.66 & 23.72 & 15.51 &   H   & B   &  	  & VPC        \\
1663  & dE & - & 17.45  & 0.71 & 25.22 & 17.33  & 2.35 & 25.86 & 23.32 &   H   & B   &  	  &	       \\
1664  &  E & - & 11.95  & 1.92 & 17.54 & 11.90  & 6.10 & 20.30 & 24.75 &   H   & B   &   5	  & F97,ACS    \\
1689  & dE & - & 16.95  & 0.79 & 24.23 & 17.46  & 2.51 & 25.07 & 14.62 &   H   & B   &  	  &	       \\
1710  & dE & - & 17.75  & 0.90 & 22.54 & 17.57  & 2.68 & 23.41 &  6.97 &   H   & B   &  	  &	       \\
1711  & dE & N & 16.43  & 0.93 & 22.47 & 16.38  & 2.81 & 23.55 & 11.95 &   H   & B   &  	  & VPC,BJ98   \\
1717  & dE & - & 16.45  & 0.57 & 24.08 & 16.91  & 2.15 & 24.60 & 22.97 &   H   & B   &  	  & BJ98       \\
1718  & dE & N & 18.25  & 1.05 & 23.96 & 18.48  & 2.86 & 25.13 & 10.67 &   H   & B   &  	  &	       \\
1729  & dE & - & 17.75  & 0.68 & 24.39 & 18.63  & 2.73 & 24.92 & 10.19 &   H   & B   &  	  &	       \\
1754  & dE & - & 18.95  & 0.45 & 25.47 & 19.42  & 1.98 & 25.87 &  8.73 &   H   & B   &  	  &	       \\
1783  & dE & N & 18.17  & 0.91 & 24.32 & 17.78  & 2.80 & 25.28 & 17.75 &   H   & B   &  	  &	       \\
1785  & dE & N & 17.75  & 0.89 & 24.00 & 17.93  & 2.59 & 24.93 & 11.11 &   H   & B   &  	  &	       \\
1794  & dE & N & 17.25  & 0.83 & 23.08 & 17.94  & 2.74 & 23.77 &  8.48 &   H   & B   &  	  &	       \\
1803  & dE & N & 16.65  & 1.14 & 22.36 & 16.23  & 2.94 & 23.73 & 12.86 &   H   & B   &  	  &	       \\
1812  & dE & N & 17.73  & 0.98 & 22.77 & 17.63  & 3.02 & 23.69 &  7.90 &   H   & B   &  	  & BJ98       \\
1814  & dE & - & 18.65  & 0.65 & 23.33 & 20.88  & 2.30 & 23.91 &  1.82 &   H   & B   &  	  &	       \\
1815  & dE & - & 17.31  & 0.49 & 24.95 & 17.92  & 1.92 & 25.08 & 13.73 &   H   & B   &  	  &	       \\
1831  & dE & N & 17.95  & 1.08 & 24.41 & 17.94  & 3.10 & 25.72 & 15.25 &   H   & B   &  	  &	       \\
1861  & dE & N & 14.31  & 2.56 & 18.83 & 14.46  & 3.14 & 22.29 & 15.04 &   H   & B   &   3	  & ACS        \\
1863  & dE & - & 18.94  & 0.54 & 24.51 & 18.75  & 2.15 & 24.87 &  9.07 &   H   & B   &  	  &	       \\
1870  & dE & - & 15.73  & 2.94 & 19.07 & 16.22  & 3.39 & 22.92 & 14.62 &   H   & B   &   3	  &	       \\
1871  &  E & - & 13.80  & 1.61 & 18.31 & 14.43  & 3.41 & 20.67 &  7.17 &   H   & B   &  	  & ACS        \\
1879  & dE & N & 17.25  & 0.95 & 23.78 & 17.32  & 2.79 & 24.90 & 14.64 &   H   & B   &  	  &	       \\
1880  & dE & - & 18.48  & 0.75 & 24.04 & 18.51  & 2.66 & 24.77 &  8.79 &   H   & B   &  	  &	       \\
1891  & dE & N & 16.63  & 1.16 & 22.37 & 17.16  & 3.31 & 23.72 & 10.41 &   H   & B   &  	  &	       \\
1901  & dE & - & 17.55  & 1.41 & 22.03 & 17.44  & 3.15 & 23.59 &  9.56 &   H   & B   &  	  &	       \\
1903  &  E & - & 10.70  & 3.03 & 15.80 & 10.88  & 5.53 & 20.60 & 39.47 &   H   & B   &   1,5	  & F97,ACS    \\
1904  & dE & N & 18.95  & 0.60 & 26.17 & 17.72  & 2.66 & 26.50 & 22.86 &   H   & B   &  	  &	       \\
1909  & dE & N & 16.05  & 1.04 & 22.35 & 16.50  & 3.22 & 23.39 & 12.78 &   H   & B   &  	  &	       \\
1910  & dE & N & 14.12  & 1.47 & 19.79 & 14.41  & 3.13 & 21.86 & 13.22 &   H   & B   &  	  & ACS        \\
1915  & dE & - & 17.08  & 0.68 & 23.89 & 17.20  & 2.65 & 24.42 & 14.58 &   H   & B   &  	  & BJ98       \\
1942  & dE & N & 16.72  & 0.79 & 23.03 & 16.81  & 2.85 & 23.96 & 13.33 &   H   & B   &  	  & BJ98       \\
1945  & dE & N & 14.77  & 1.01 & 22.00 & 15.23  & 3.48 & 23.15 & 21.61 &   H   & B   &  	  &	       \\
1951  & dE & N & 16.95  & 1.16 & 23.66 & 16.92  & 3.14 & 24.91 & 16.10 &   H   & B   &  	  &	       \\
1958  & dE & N & 16.93  & 0.92 & 22.57 & 16.88  & 2.86 & 23.52 &  9.45 &   H   & B   &  	  &	       \\
1971  & dE & - & 16.53  & 1.19 & 22.02 & 16.72  & 3.24 & 23.35 & 10.78 &   H   & B   &  	  &	       \\
1982  & dE & - & 15.24  & 0.87 & 21.84 & 16.02  & 2.72 & 22.84 & 13.67 &   H   & B   &   1	  &	       \\
1986  & dE & N & 18.95  & 1.35 & 24.68 & 18.16  & 3.55 & 26.11 & 16.00 &   H   & B   &  	  &	       \\
1991  & dE & N & 15.55  & 1.12 & 22.54 & 15.66  & 2.91 & 23.93 & 20.13 &   H   & B   &  	  &	       \\
1995  & dE & - & 15.75  & 1.05 & 22.64 & 15.90  & 2.81 & 23.95 & 18.18 &   H   & B   &  	  &	       \\
2000  &  E & - & 11.87  & 2.17 & 16.03 & 12.11  & 4.84 & 19.46 & 13.56 &   H   & B   &   3,5	  & ACS*       \\
2001  & dE & - & 18.95  & 0.79 & 24.26 & 18.02  & 2.54 & 25.16 & 11.46 &   H   & B   &  	  &	       \\
2003  & dE & - & 18.15  & 0.96 & 24.33 & 17.98  & 3.06 & 25.34 & 15.50 &   H   & B   &  	  &	       \\
2008  & dE & - & 14.95  & 0.86 & 23.16 & 15.06  & 2.72 & 24.21 & 40.74 &   H   & B   &  	  &	       \\
2010  & dE & - & 18.95  & 0.69 & 24.42 & 18.88  & 2.39 & 25.07 &  8.35 &   H   & B   &  	  &	       \\
2011  & dE & - & 17.05  & 0.72 & 24.00 & 17.40  & 2.61 & 24.65 & 13.05 &   H   & B   &  	  &	       \\
2012  & dE & N & 14.22  & 1.30 & 21.77 & 14.74  & 2.97 & 23.30 & 22.42 &   H   & B   &   1	  &	       \\
2025  & dE & - & 18.45  & 1.32 & 23.80 & 18.34  & 3.83 & 25.62 & 13.80 &   H   & B   &  	  &	       \\
2032  & dE & - & 17.45  & 1.08 & 24.60 & 17.18  & 2.95 & 25.91 & 24.12 &   H   & B   &  	  &	       \\
2049  & dE & N & 16.35  & 0.65 & 22.72 & 16.52  & 2.38 & 23.26 & 15.19 &   H   & B   &   1	  &	       \\
2050  & dE & N & 15.13  & 1.12 & 21.05 & 15.23  & 3.03 & 22.98 & 16.37 &   H   & B   &  	  & ACS        \\
2051  & dE & - & 17.45  & 0.78 & 22.91 & 17.24  & 2.65 & 23.75 &  9.29 &   H   & B   &  	  &	       \\
2056  & dE & - & 16.95  & 1.15 & 21.53 & 16.43  & 3.18 & 22.98 &  9.59 &   H   & B   &  	  &	       \\
2072  & dE & - & 18.95  & 0.47 & 25.10 & 19.54  & 1.76 & 25.53 &  7.31 &   H   & B   &  	  &	       \\
2078  & dE & - & 17.45  & 1.15 & 24.10 & 16.26  & 3.24 & 25.49 & 32.61 &   H   & B   &  	  &	       \\
2081  & dE & - & 16.95  & 0.82 & 23.59 & 17.11  & 2.70 & 24.47 & 13.82 &   H   & B   &  	  &	       \\
\noalign{\smallskip}
\hline
\multicolumn{14}{l}{Column 1: VCC designation.}          \\
\multicolumn{14}{l}{Column 2,3: Morphological type from the VCC. N=Nucleated.}          \\
\multicolumn{14}{l}{Column 4: Photographic magnitude $m_p$ from the VCC.}\\
\multicolumn{14}{l}{Column 5: Sersic index $n$.}\\
\multicolumn{14}{l}{Column 6: Sersic extrapolation to $r=0$ of the central surface brightness $\mu_o$ (in $\rm mag~arcsec^{-2}$).}\\
\multicolumn{14}{l}{Column 7: Total asymptotic magnitude $B_T$.}\\ 
\multicolumn{14}{l}{Column 8: Light concentration index $C_{31}$.}\\ 
\multicolumn{14}{l}{Column 9: Mean effective surface brightness $<\mu>_e$ (in $\rm mag~arcsec^{-2}$).}\\
\multicolumn{14}{l}{Column 10: Effective major-axis radius $r_e$ (in arcsec).}\\
\multicolumn{14}{l}{Column 11: WFS strip: H=horizontal; V=vertical.}\\
\multicolumn{14}{l}{Column 12: Available band(s).}\\
\multicolumn{14}{l}{Column 13: Notes. (1 = poor outer fit; 2 = M32 like object; 3 = two seeing disks excluded from the fit; 4 = core;}\\ 
\multicolumn{14}{l}{~~~~~~~~~~~~~~5 = power law; 6 = saturated.}\\
\multicolumn{14}{l}{Column 14: Reference to other works. B03 = Barazza et al (2003), F97 = Faber et al. (1997), VPC = Young \& Currie (1998),}\\
\multicolumn{14}{l}{~~~~~~~~~~~~~~BJ98 = Binggeli \& Jerijn 1998; ACS = in The ACS Virgo Cluster Survey,}\\
\multicolumn{14}{l}{~~~~~~~~~~~~~~ASC*= whether core or power law, kindly provided by Laura Ferrarese (private communication).}\\
\end{longtable}
  \clearpage										  

   \clearpage

\end{document}